\documentclass{aa}  

\usepackage{txfonts}
\usepackage[export]{adjustbox}
\usepackage{subcaption}
\usepackage[colorlinks,linkcolor=blue,anchorcolor=blue,citecolor=blue]{hyperref}
\usepackage{xcolor}
\newcommand{\usfr}{$\rm M_{\sun}\,yr^{-1}$}
\newcommand{\usfrd}{$\rm M_{\sun}\,yr^{-1}\,kpc^{-2}$}

\newcommand{\micron}{$\muup\rm m$}
\newcommand{\ulum}{$\rm W\,Hz^{-1}$}
\newcommand{\uvel}{$\rm km\,s^{-1}$}

\begin{document} 

    \title{Nearby galaxies in the LOFAR Two-metre Sky Survey}

   \subtitle{IV. A fundamental plane of the radio–SFR relation}
   \author{V. Heesen\inst{1}
   \and
   H.~W.~Edler\inst{2}
   \and
   M.~Br\"uggen\inst{1}
   \and
   M.~Stein\inst{3}
   \and
   D.~J.~Bomans\inst{3}
   \and
   R.~Paladino\inst{4,5}
   \and
   K.~T.~Chy\.zy\inst{6}
   \and
   K.~Ma{\l}ek\inst{7,8}
   \and
   M.~A.~Lara-L\'opez\inst{9,10}
   \and
   F.~S.~Tabatabaei\inst{11}
   }

   \institute{Hamburger Sternwarte, University of Hamburg, Gojenbergsweg 112, 
21029 Hamburg, Germany\\
    \email{volker.heesen@uni-hamburg.de}
    \and
    ASTRON, the Netherlands Institute for Radio Astronomy, Postbus 2,
NL-7990 AA Dwingeloo, The Netherlands
    \and
    Ruhr University Bochum, Faculty of Physics and Astronomy, Astronomical Institute (AIRUB), 44780 Bochum, Germany
    \and
    European Southern Observatory, Karl-Schwarzschild-Strasse 2, 85748 Garching, Germany
    \and
    INAF-Istituto di Radioastronomia, Via P. Gobetti 101, 40129 Bologna,
Italy
    \and 
    Astronomical Observatory of the Jagiellonian University, ul. Orla 171, 30-244 Krak\'ow, Poland
    \and
    National Centre for Nuclear Research, Pasteura 7, 02-093, Warsaw, Poland
    \and
    Aix-Marseille Université, CNRS, CNES, LAM, Marseille, France
    \and
    Departamento de Física de la Tierra Astrofísica. Universidad Complutense de Madrid (UCM), E-28040 Madrid, Spain
    \and
    Instituto de Física de Partículas y del Cosmos, IPARCOS, Fac. C.C. Físicas, Universidad Complutense de Madrid, E-28040 Madrid, Spain
    \and
    School of Astronomy, Institute for Research in Fundamental Sciences (IPM), P.O. Box 1956836613, Tehran, Iran
    }

   \date{Received date / Accepted date}

 
  \abstract
   {Radio continuum emission has the potential to be an extinction-free tracer of star formation. However, the relation between radio continuum luminosity and star formation rate, the radio-SFR relation, is potentially limited by various effects such as cosmic-ray transport, free-free absorption, and cosmic-ray electron energy losses.}
   {We aim to calibrate the radio-SFR relation in a sample of nearby galaxies ranging from dwarf  to  spiral galaxies covering nearly five orders of magnitude in SFR range. We include, both, global (individual galaxies) and local (spatially resolved, kiloparsec sized) measurements.}
   {We measured radio continuum luminosities at 144\,MHz using observations with the LOw Frequency ARray (LOFAR) and measure radio spectral indices using ancillary 1.4\,GHz data. Selecting 70 nearby (distance $d<30\,\rm Mpc$) galaxies, 15 of which were used for local measurements, with rich ancillary data we present a study of the radio-SFR relation using total infrared, mid-infrared,  H$\alpha$, and far-ultraviolet as complementary SFR tracers. About one third of our sample are at least moderately star-forming edge-on galaxies with the remaining ones chosen to be a representative sample of a wide range of morphological types and SFR values.}
   {For the first time, we show that the radio luminosity ($L_{144}$), the star-formation rate (SFR), and the radio spectral index ($\alpha$) define a `fundamental plane' in the [$\log(L_{144})$, $\log({\rm SFR})$, $\alpha$] space. This allows us to define a unified radio-SFR relation that works both for global and local data when using the radio spectral index as a second parameter.}
   {A unified radio-SFR relation for, both, global and local data may serve as a litmus test for galaxy simulations that include the effect of cosmic rays and magnetic fields. It also strengthens the case for using the radio-SFR relation as an extinction-free tracer of star formation.}




\titlerunning{A fundamental plane of the radio--SFR relation}
\authorrunning{V.~Heesen et al.}

   \maketitle
%

\section{Introduction}

The star-formation rate (SFR) is one of the fundamental parameters to probe galaxy formation and evolution. Hence, considerable effort has gone into the calibration of various star formation tracers across the electromagnetic spectrum using, both, line and continuum emission \citep{kennicutt_12a}. Radio continuum is one of those tracers \citep{murphy_11a}, and it offers us some unique benefits such as high angular resolution using interferometric observations, extinction-free star formation measurements, ground-based observations, and wide-field survey capabilities. Also, radio continuum is one of the few avenues to study both cosmic rays and magnetic fields in galaxies, some of the crucial ingredients for state-of-the-art galaxy evolution models \citep{thompson_24a}.  Recent developments in high-performance computing made it possible to conduct all-sky surveys with radio interferometers at high angular resolution and good sensitivity such as the Evolutionary map of the Universe \citep[EMU;][]{norris_21a} survey with the Australian Square Kilometre Array Pathfinder (ASKAP) and the LOFAR Two-metre Sky Survey \citep{shimwell_17a, shimwell_19a} and \citet{shimwell_26a} with the LOw Frequency ARray \citep[LOFAR;][]{vanHaarlem_13a}.

Analysis of the LOFAR data has shown that galaxies follow a close relation between radio continuum luminosity and SFR, the radio-SFR relation in short \citep[e.g.][]{gurkan_18a}. At these low frequencies the relation is significantly affected not just by the SFR but also by the mass: galaxies that are more massive, are also brighter in the radio continuum compared to their SFR \citep{smith_21a,shenoy_26a}. One possible explanation is that more massive galaxies are better cosmic-ray electron (CRE) calorimeters and thus have steeper radio continuum spectra \citep{heesen_22a, edler24}. There are several processes that shape the radio continuum spectra such as free-free absorption and cosmic-ray ionisation losses that suppress the emission at low frequencies and thus flatten the spectrum \citep{basu_15a}. On the other hand, synchrotron and inverse Compton radiation losses steepen the radio continuum spectrum. Hence, there is a complex interplay with cosmic-ray ageing, transport, and acceleration that governs the details of the radio-SFR relation \citep{roth_23a}. 

These processes can explain why the relation is strongly super-linear for global measurements \citep[e.g.][]{bell_03a,li_16a,heesen_22a}, while it is strongly sub-linear for local measurements \citep[e.g.][]{basu_12a,heesen_24a}. In order to improve the radio-SFR calibration, it was shown in earlier work that restricting the analysis to areas near star formation sites can reduce the scatter significantly \citep{dumas_11a,basu_12a}. In later work it was shown that the radio spectral index can be a good proxy for these areas and  selecting only data in agreement with the CRE injection spectrum \citep{heesen_14a,heesen_19a}. As an extension to these papers it was shown that taking into account the radio continuum spectrum, the radio-SFR relation can be linearised using the radio spectral index as an additional information \citep{heesen_24a} or using more than one frequency \citep{tabatabaei_17a}. One aspect that is particularly exciting is the prospect of a unified radio-SFR relation that holds, both, at global and local i.e.\ kiloparsec scales \citep{heesen_24a}.

This empirical calibration showed a strong relation between the relative radio brightness, i.e. the ratio of radio intensity to SFR surface density, as function of radio spectral index \citep{heesen_24a}. It was also shown that this relation can be applied to global measurements in order to approximately linearise the radio-SFR relation. Qualitatively this can be understood as a superposition of processes that flatten the spectrum and thus suppress radio continuum emission at low frequencies, such as cosmic-ray ionisation losses and free-free absorption. On the other hand synchrotron radiation losses steepen the spectrum and result in radio-bright galaxies \citep{lacki_10a}. Processes that flatten the spectrum have been observed in nearby starburst galaxies where they affect the global spectrum \citep{adebahr_13a,kapinska_17a} or in spatially resolved observations of nearby galaxies \citep{gajovic_24a,gajovic_25a}. Simulations indicate that galaxies might be good electron calorimeters and a low-frequency flattening stems from free-free absorption \citep{werhahn_21c}. 

In this work we wish to explore the low-frequency radio-SFR relation both on a global and local (kiloparsec sized) basis. We use images from the third data release of LoTSS \citep[LoTSS-DR3;][]{shimwell_26a} to probe galaxies from low-mass dwarf galaxies to massive spiral galaxies. We then relate the ratio of radio to ancillary SFR surface brightness to the radio spectral index and show that the global and local data indeed follow the same trend. This potentially allows us to define a unique radio-SFR relation where the radio spectral index is used as a second parameter. This work albeit only expanding the sample of LoTSS-DR2 \citep{heesen_22a} by about 50 per cent, extends the mass and SFR range to now four orders of magnitude (previously less then three orders of magnitude). Extending our previous work in \citet{heesen_24a} where we analysed the local measurements we now expand this work to a consistent description of both global and local data making this distinction obsolete.

This paper is organised as follows: Section\,\ref{s:data_and_methodology} describes our sample selection, data reduction, and methodology. In Sect.\,\ref{s:results} we present our resulting radio-SFR relations and radio spectral indices that we relate to fundamental galaxy parameters. We discuss our results in Sect.\,\ref{s:discussion} before we conclude in Sect.\,\ref{s:conclusions}. Appendix\,\ref{as:results} contains our presented observational results.

\section{Data and methodology}
\label{s:data_and_methodology}

\subsection{Galaxy sample}
\label{ss:galaxy_sample}

We defined our galaxy sample in \citet{heesen_22a}. In brief, we draw them from the SIRTF Nearby Galaxies Survey \citep[SINGS;][]{kennicutt_03a}, Key Insights on Nearby Galaxies: A Far-Infrared Survey with Herschel: Survey Description and Image Atlas \citep[KINGFISH;][]{kennicutt_11a} and Continuum Halos in nearby galaxies: An EVLA survey \citep[CHANG-ES;][]{irwin_12a,irwin_24a}. The three galaxies 
which are only in KINGFISH but not in SINGS are NGC\,2146, IC\, 342, and NGC 5457. Finally, we added another four galaxies, namely IC\,10, NGC\,598 (M\,33), NGC\,4214, and NGC\,4449. Three of these are
star-burst dwarf irregular galaxies and M\,33 is the most nearby galaxy, which we can easily image as it is less than $1\degr$ in apparent size. The SINGS and Kingfish samples contain nearby galaxies (distance $d<30\,\rm Mpc$), and are chosen in such a way that they are representative in morphological type, infrared luminosity (proxy for SFR) and far-infrared-to-optical luminosity ratio (proxy for specific SFR). The CHANG-ES galaxies are chosen as edge-on galaxies (inclination angle $i>75\degr$), with an optical size between $4$ and $15\arcmin$ and a radio flux density of $>$23\,mJy at $1.4\,\rm GHz$.

In total, the sample contains 76 galaxies, 45 of which we published already in \citet{heesen_22a} based on LoTSS-DR2 data. With LoTSS-DR3, we were able to increase the sample size and so to include a number of low-mass dwarf galaxies which extends our sample to galaxies with only little star formation. Of the 76 galaxies, we found 70 galaxies suitable for our analysis. We left out six galaxies, namely M81\,DwA, Ho\,IX, NGC\,3190, NGC\,4536, DDO\,154, and DDO\,165. The dwarf galaxies M81\, DwA, Ho\,IX, DDO\,154, and DDO\,165 were non-detections. We searched at their position for a $2\sigma$ detection in the $20\arcsec$ maps with no detection. We discarded in those cases point-like sources at $6\arcsec$, likely background galaxies. These were checked with Pan-STARRS \citep[Panoramic Survey Telescope and Rapid Response System;][]{chambers_16a} optical overlays. NGC\,3190 is contaminated by a background source, a Fanaroff--Riley type 2 radio galaxy. NGC\,4536 was left out as the image quality was far inferior compared with the other data. This is probably caused by a combination of low declination and proximity to Virgo\,A.

We present our data in these four different sub-groups of CHANG-ES (21), SINGS (42), Kingfish+ (three), and `Extra' (four) galaxies. These individual galaxies are referred to as `global' data in what follows. We complement the global data with spatially resolved data. For this we use the galaxy sample as presented in \citet{heesen_24a}. These include 15 galaxies from the SINGS sample. For these galaxies a spatial resolution of approximately one kiloparsec is reached. These measurements are referred to as `local' data.

\subsection{LOFAR data}
\label{ss:lofar_data}

We use data from the third data release of the LOFAR Two-metre Sky Survey \citep[LoTSS-DR3;][]{shimwell_26a} in conjunction with previously analysed data from the second data release \citep{heesen_22a,shimwell_22a}. We use the maps with $20\arcsec$ angular resolution; alternatively, there are also maps at $6\arcsec$ angular resolution.\footnote{\href{https://lofar-surveys.org}{https://lofar-surveys.org}} For some galaxies (NGC\,4254, 4321, and 4450) that are in the vicinity of Virgo\,A (M\,87), we used data from the `VIrgo Cluster multi-Telescope Observations in Radio of Interacting galaxies and AGN' project \citep[VICTORIA;][]{edler_23a}. These data are taken in an identical setup to LoTSS but are especially calibrated and imaged maps that can deal with the high-dynamic range required. The map of IC\,10 was obtained from \citet{heesen_18a} and has a lower angular resolution of $44\arcsec$. We then defined elliptical contours based on the optical centre (obtained from the NASA Extraglactic Data Base) and position angle (see Appendix~\ref{as:results}) such that they approximately enclose the the $3\sigma$ contours of the 144\,MHz radio continuum map. The intensity is then integrated within these elliptical regions to obtain the flux density. In a few cases (NGC\,4236 and 4254) we subtracted the flux density of unrelated background sources, but in most cases they could be neglected within the uncertainties. 


The radio flux densities were converted into radio luminosities without $k$-correction, meaning we assume that the rest frequency is equivalent to the observed frequency. As the redshift of our sample galaxies is $<0.01$, this is a good approximation. Hence, we obtain for the radio continuum luminosity at 144\,MHz:
\begin{equation}
 L_{144}=4\piup  d^2 S_\nu,
 \label{eq:luminosity}
\end{equation}
where $S_\nu$ is the integrated flux density at 144\,MHz and $d$ is the distance. 

\subsection{Star formation rates}
\label{ss:star_formation_rates}

The SFRs were calculated in two different ways (see also Appendix\,\ref{as:star_formation_rates}). First, for global data the we used a combination of H$\alpha$ and mid-infrared observations to calculate dust-corrected H$\alpha$ luminosities. Second, for the local data, we used a combination of mid-infrared and far-ultraviolet observations. While it would be preferable of course to have an entirely consistent approach, these two methods are consistent with each other to an accuracy of about 0.1\,dex at SFR values $>10^{-3}$\,\usfr, where the bulk of our data points reside \citep{leroy_12a}. At lower SFRs, the uncertainties become larger of course. The limitation of our approach is that for galaxies at higher masses the mid-infrared emission becomes optically thick so that the SFRs are systematically too low \citep{vargas_19a}. The alternative is to use total infrared (TIR) luminosities to measure SFRs. However, at low masses, the TIR luminosities are too low since dwarf galaxies are dust deficient. In Appendix\,\ref{as:tir_results} we show selected alternative results when using the TIR SFRs for the global data points. In Appendix\,\ref{as:comparison_of_sfrs} we show a comparison of the different SFRs and further motivate our choice by showing that the global and local SFRs are indeed consistent with our approach.

Hence, for the global measurements, we used dust-corrected H$\alpha$ luminosities. These were adopted from \citet{kennicutt_11a} based on the calibration presented in \citet{kennicutt_09a} and tabulated in \citet{calzetti_10a}. The method combines H$\alpha$ data with mid-infrared data at 24\,\micron. For the CHANG-ES sample we simply used the SFRs from \citet{wiegert_15a}. They are solely based on the 22\,\micron\ fluxes with the \emph{WISE} satellite. These are then converted using the relation presented in \citet{jarrett_13a} \citep[see also][]{rieke_09a}. This assumes that the
mid-infrared emission captures all of the star-forming activity and that we do not need additional ultraviolet or H$\alpha$ observations to estimate the SFR. Since the CHANG-ES galaxies are highly inclined thus at high optical depth, this should be a reasonable assumption. The more likely source of bias, affecting all our SFRs, is including `diffuse' mid-infrared emission that reflects dust heated by an older stellar population \citep[e.g.][]{belfiore_23a}. For any galaxies not in these samples, we used archival H$\alpha$ and \emph{Spitzer} 24\,\micron\ data to calculate dust-corrected H$\alpha$ luminosities. These are then converted into the SFRs using the conversion by \citet{kennicutt_09a}.

For the local SFR measurements, we used the SFR surface density maps of \citet{leroy_08a} which are based on a combination of \emph{GALEX} far-ultraviolet data at 156\,mn and \emph{Spitzer} 24\,\micron\ data. These are excellent maps with good angular resolution of $13\farcs 5$ and very good sensitivity of $<10^{-4}$\,\usfrd.

\subsection{Local data}
\label{ss:local_data}

In order to generate the local data, we used the software {\tt radio-pixel-plots} ({\sc rpp}).\footnote{\href{https://github.com/sebastian-schulz/radio-pixel-plots}{https://github.com/sebastian-schulz/radio-pixel-plots}} For this we used radio continuum maps at 144\,MHz from LOFAR and at 1365\,MHz from the Westerbork Synthesis Radio Telescope \citep[WSRT;][]{braun_07a}. We also used hybrid SFR surface density maps based on a combination of \emph{GALEX} far-ultraviolet data at 156\,mn and \emph{Spitzer} 24\,\micron\ data \citep{leroy_08a}.

To compare the global SFRs with those from spatially resolved measurements, we converted the radio intensities into luminosities. Basically, we convolved the maps with a Gaussian to an angular resolution where the full width at half maximum corresponds to a projected distance of 1.2\,kpc (for some galaxies we chose slightly larger values). Then the maps were binned into pixel elements with 1.2\,kpc side length. These values were then extracted using a 3$\sigma$ lower cutoff in radio continuum emission at 144 and 1365\,MHz and in SFR surface density \citep[see][for more details]{heesen_24a}.\footnote{Since the output of {\sc rpp} is the radio SFR surface density converted according to Eq.\,(3) in \citet{heesen_24a}, we first converted them back to intensities.} In order to calculate radio continuum luminosities we used: 
\begin{equation}
    L_{144}=4\piup \Omega d^2 I_\nu,
    \label{eq:luminosity_resolved}
\end{equation}
where $\Omega$ is the solid angle of the area $A$ observed in the galaxy. Our maps were convolved to a Gaussian beam such that $\Omega=2\sqrt{2\ln 2}\rm FWHM^2$, where $\rm FWHM$ is the full width at half maximum of the Gaussian beam. The FWHM is chosen such that the projected length in the galaxy is equivalent to 1.2\,kpc.

 The star formation rates were calculated with ${\rm SFR}=A\Sigma_{\rm SFR}$, where $A=1.2\times1.2\,\rm kpc^2$ (in some cases the area is slightly larger). In total, there are 2119 local data points In detail, NGC\,628 (132 data points), NGC\,925 (176), NGC\,2403 (68), NGC\,2841 (190), NGC\,2903 (98), NGC\,2976 (17), NGC\,3184 (238), NGC\,3198 (53), NGC\,3938 (170), NGC\,4254 (233), NGC\,4725 (103), NGC\,4736 (46), NGC\,5055 (146), NGC\,5194 and NGC\,5195 (278), and NGC\,7331 (171). Of course, we would expect that the sum of the local SFRs is similar to the global SFR for each galaxy for consistency. This is indeed at least approximately the case (see Appendix~\ref{as:comparison_of_sfrs}). Small deviaions are expected because the local data exclude small parts of the galaxies in particular galactic nuclei which may be influenced by an active galactic nucleus and some prominent background sources. These parts of the SFR maps were masked.

\subsection{Radio spectral indices}
\label{ss:radio_spectral_indices}

We obtained flux densities at around $1.4$\,GHz from the literature in order to calculate radio spectral indices. To this end we use mostly data at 1365\,MHz from the Westerbork Synthesis Radio Telescope \citep[WSRT;][]{braun_07a} and data at 1570\,MHz from the Jansky Very Large Array \citep[JVLA;][]{wiegert_15a}. For those galaxies where they were not available we use other data from the literature such as from the 100\,m Effelsberg telescope \citep{tabatabaei_17a} and the Jansky Very Large Array \citep[eg.][]{yun_01a}. Details can be found in Appendix\,\ref{as:radio_spectral_indices}. Radio spectral indices are then calculated following the convention $S_\nu\propto \nu^\alpha$. 

In a few cases (IC\,2574 and DDO\,53) instead of using the published flux densities, we re-measured the flux densities directly from the published maps using our own apertures. The reason is that our assumed apertures were very different from those in the literature. The reason is that the LOFAR maps show more extended emission, hence our integration regions were larger. In NGC\,4236 and 4254 we measured the flux densities with our own regions and also subtracting background sources as we did for the LOFAR data.

\subsection{Uncertainties, binning, and fitting of data}
\label{ss:uncertainty_estimates}

For the LOFAR data, we assumed an absolute flux-density scale uncertainty of 10\,\%, following \citet{shimwell_26a}; for the galaxies from the VICTORIA survey with LOFAR we assumed 20\,\% uncertainty \citep{edler_23a}. For the global and local SFR measurements, we assume an uncertainty of 8\,\%. This value should be considered as flux uncertainty only, assuming 15\,\% uncertainty of either the H$\alpha$  \citep{kennicutt_09a} or the far-ultraviolet flux \citep{gil_de_paz_07a} and 2\,\% for the mid-infrared flux \citep{jarrett_13a}. This assumes that both are contributing about the same to the SFR \citep{leroy_12a}. We note that the actual calibration of SFRs even for global SFR measurements are about only 50\,\% accurate at best; for local measurements the uncertainties are even larger \citep{leroy_12a}. Local measurements in the radio continuum may have also a larger uncertainty, in particular in areas of low signal-to-noise ratios. This we have attempted to mitigate using a 3\,$\sigma$ lower cut-off in radio continuum intensities, where $\sigma$ the RMS noise value of the intensity map.

We binned our local data into twelve bins, separated by 0.25\,dex in SFR values. For these we calculated the mean of the logarithmic radio luminosities and the mean of the logarithmic SFR values as well as the mean radio spectral index. For uncertainties we took the standard deviation of the mean. These data are referred to as `binned local data'. We fit our data to both global and binned local data, unless otherwise mentioned. This in effect applies a weighting to the fitting such that we prevent giving a lot of weight to the local measurements since there are so many more of them.

Best-fitting $f(x)$ relations were obtained with the orthogonal distance regression ({\sc odr}) method. This method can take $x$ uncertainties into account. In our case, this was done where $x$ is either the SFR or the radio spectral index.  For best-fitting $f(x, y)$ bivariate relations with two variables we took only errors of the fitted values into account.

\begin{figure}
    \includegraphics[width=\linewidth,valign=t]{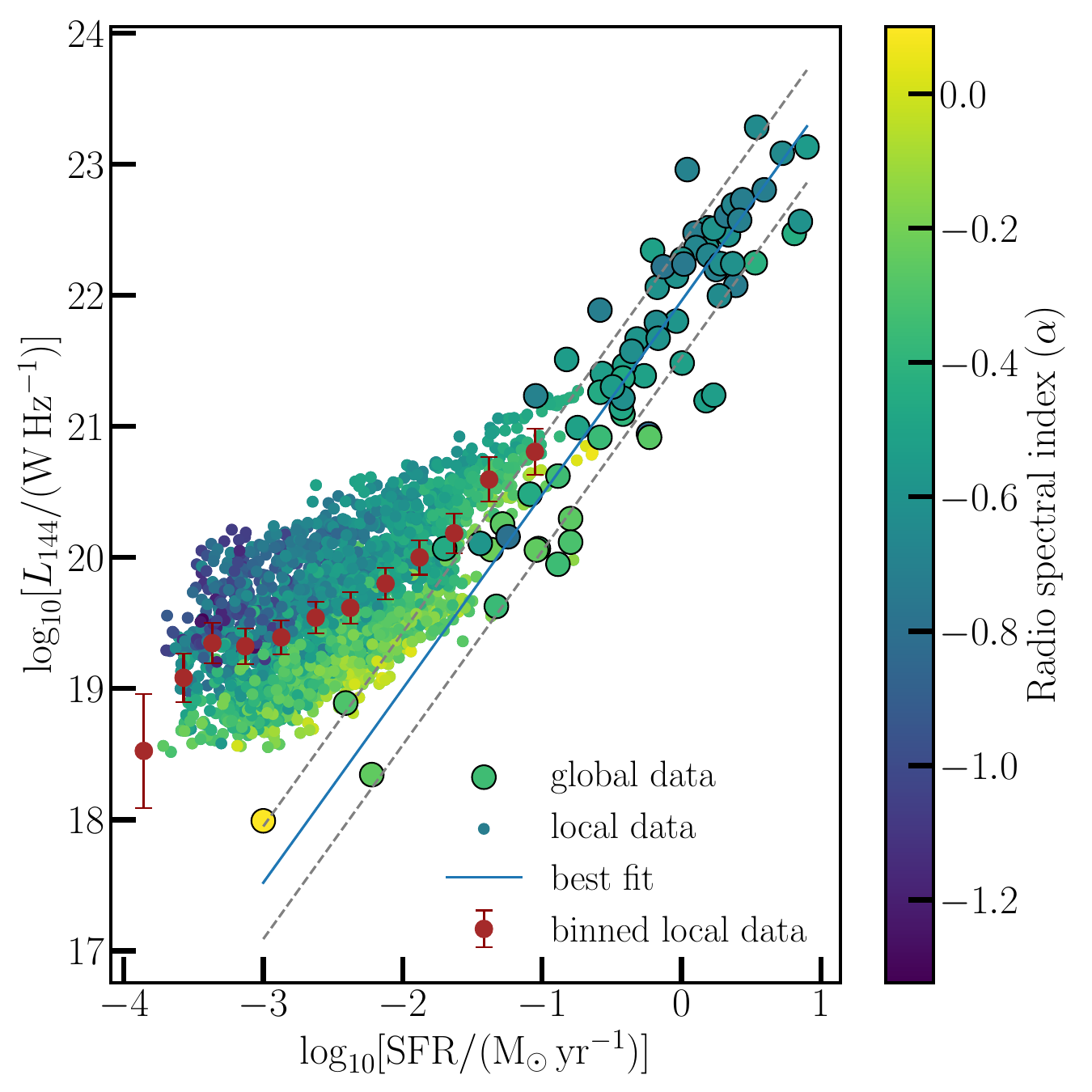}
    \caption{Radio-SFR relation. Radio continuum luminosity at 144\,MHz as a function of the SFR. Large data points show global and small data points local measurements (the latter largely overlapping). The best-fitting radio-SFR relation to the global data is shown as well (solid line with a $\pm 1\,\sigma_y$ interval indicated by dashed lines). Red data points with error bars show binned local data. The SFR was derived from a combination of mid-infrared data with either H$\alpha$ (global) or far-ultraviolet (local data). Data points are coloured according to their radio spectral index between 144 \,MHz and approximately 1400\,MHz.}
    \label{fig:radio_sfr}
\end{figure}

\section{Results}
\label{s:results}

\subsection{Radio-SFR relation}
\label{ss:radio_sfr_relation}

First, we investigate the global radio-SFR relation. For this we parametrise the relation in the following way:
\begin{equation}
    \log_{10}(L_{144}) = a \log_{10}({\rm SFR}) + b,
    \label{eq:radio_sfr}
\end{equation}
where $L_{144}$ is the spectral luminosity at 144\,MHz in units of \ulum and SFR is the star-formation rate in units of \usfr.

Therefore, first, we investigate the global radio continuum luminosities at 144\,MHz as function of the SFR. The resulting relation is presented in Fig.\,\ref{fig:radio_sfr}. The data points fitted include only the global measurements, but for comparison we also show the local measurements. We find a best-fitting power-law as described Equation\,\eqref{eq:radio_sfr} with $a=1.48\pm 0.07$ with a RMS data scatter of $\sigma_y=0.44$\,dex (see Table\,\ref{tab:fit} for the results this and the other fits to the global data presented in this work).  We recall that for the global relation we find $a=1.4$--$1.5$ \citep{heesen_22a}, meaning a strong super-linear relation. In contrast, for the local relation we find $a=0.6$--$0.8$, meaning a strong sub-linear relation \citep{heesen_24a}. Therefore, we do not expect that we can fit both data sets simultaneously. Still, fitting both global and local data together results in a large scatter of $\sigma_y=0.67$\,dex.

We now compare our spatially resolved measurements with the global relation. These are shown as the small data points in Fig.\,\ref{fig:radio_sfr}. These points lie all on the left side of the plot at low values of the SFR. As can be seen clearly, they  do not fit to the global relation, but mostly lie above it. The deviation is a strong function of radio spectral index with data points with a steep spectrum lying furthest above the relation. Mostly, only data points with a radio spectral index indicative of a flat spectrum are in agreement with the global relation; although, even they may deviate significantly from it (in particular those at the lower-left end of the distribution). This behaviour has been interpreted in the literature as a result of cosmic-ray transport such as diffusion \citep{heesen_23a}. In our previous work we argued that energy-independent diffusion describes the data best \citep{heesen_24a}. Aside from cosmic-ray transport there may be influences of bursty star formation which can increase the scatter in the radio-SFR relation due to the different time-scales involved and the reliability of SFR estimates \citep{haskell_23a}. On the other hand, one may expect \ion{H}{ii} regions to lie below the global relation since radio emission is absorbed and CRE quickly escape these regions. We do not see such data points, possibly this can only be observed at frequencies below 100\,MHz \citep{gajovic_25a}.


\begin{table}
\caption{Best-fitting global relations using the equation $\log_{10}(y) = a\log_{10}(x) + b$ for $y=L_{144}$ or $y =   a\log_{10}(x) + b$ for $y=\alpha$. The variable $\rho$ denotes Spearman's correlation coefficient.}
\label{tab:fit}
\centering
\begin{tabular}{l c c c c}
\hline\hline 
$y$    & $a$ & $b$ & $x$ & $\rho$\\
\hline
$L_{144}$ &     $1.480\pm 0.070$ & $21.957\pm 0.057$ & $\rm SFR$ & $0.90$ \\ 
$L_{144}$ &     $1.180\pm 0.051$ & $21.57\pm 0.048$ & $\rm SFR_{\rm TIR}$   & $0.93$ \\ 
$\alpha$ &    $-0.131\pm 0.015$ & $0.76\pm 0.15$ & $M_{\star}$      & $-0.68$ \\
$\alpha$ &    $0.132\pm 0.019$ & $-0.590\pm 0.014$  & $\rm SFR$        & $-0.55$ \\
$\alpha$ &    $0.107\pm 0.016$ & $-0.553\pm 0.014$  & $\rm SFR_{\rm TIR}$        & $-0.55$ \\
$\alpha$ &    $-0.28\pm 0.03$ & $-0.28\pm 0.03$ & $r_\star$         & $-0.70$ \\
$\alpha$ &    $-0.420\pm 0.045$ & $0.34\pm 0.10$ & $v_{\rm rot}$       & $-0.67$ \\
$\alpha$ &    $0.023\pm 0.014$ & $-0.416\pm 0.089$ & $\Sigma_{\rm SFR}$  & $0.24$ \\
$\alpha$ &    $0.006\pm 0.016$ & $-0.53\pm 0.09$ & $\Sigma_{\rm SFR, TIR}$  & $0.15$ \\
\hline               
\end{tabular}
\end{table}

\begin{figure}
    \centering
    \includegraphics[width=\linewidth]{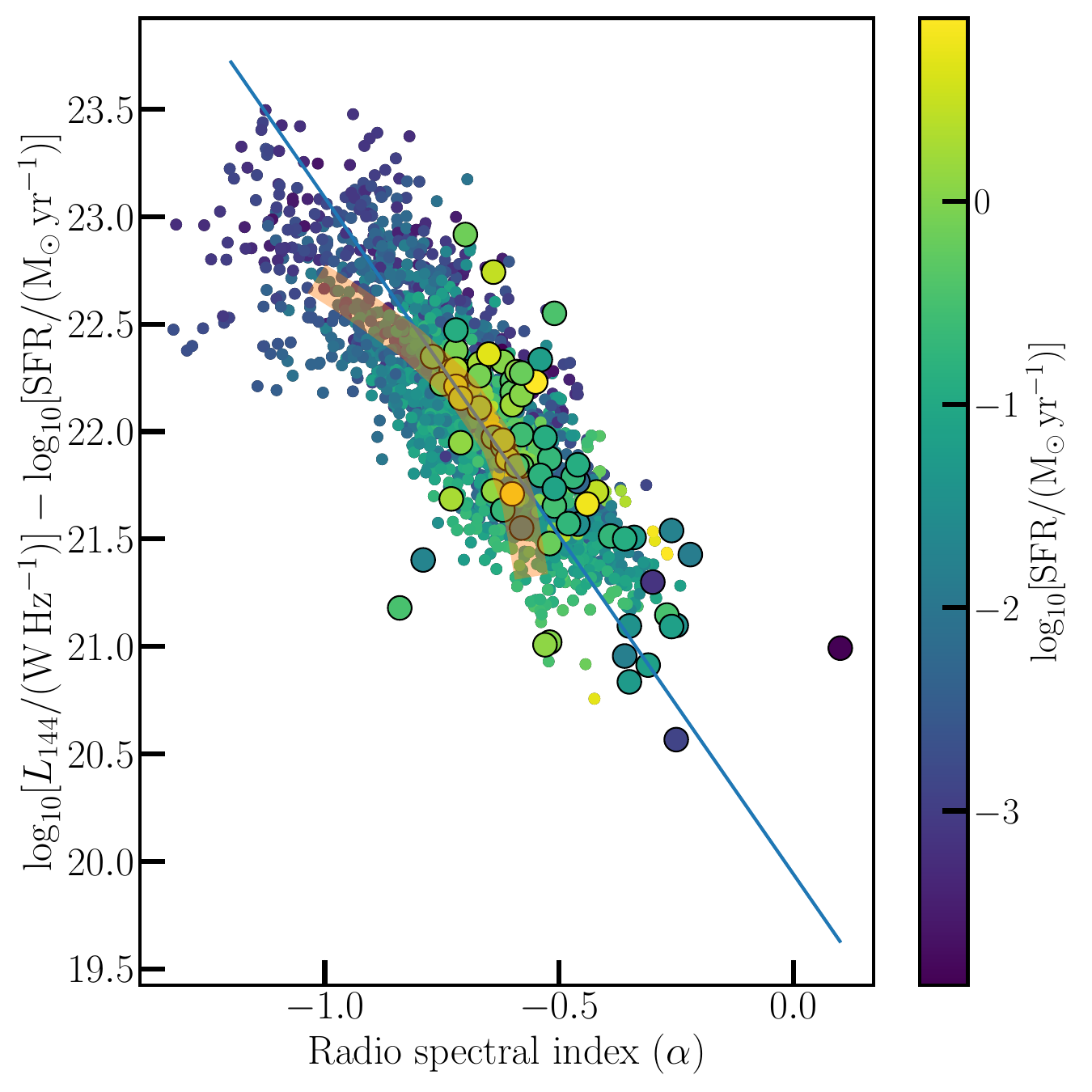}
    \caption{Radio-to-SFR ratio as function of radio spectral index. We show the ratio of radio continuum luminosity to the SFR as function of radio spectral index. Large data points show global measurements and small data points local measurements. The blue line shows the best-fitting relation. The orange shaded thick line shows a model for a semi-calorimetric galaxy with cosmic-ray electrons radiation losses (see Sect\,\ref{ss:cosmic_ray_electron_calorimetry}).}
    \label{fig:ratio_alpha}
\end{figure}

\begin{figure*}
    \centering
     \centering
     \begin{subfigure}[t]{0.03\linewidth}
        \textbf{(a)}    
    \end{subfigure}
    \begin{subfigure}[t]{0.46\linewidth}
    \includegraphics[width=\linewidth,valign=t]{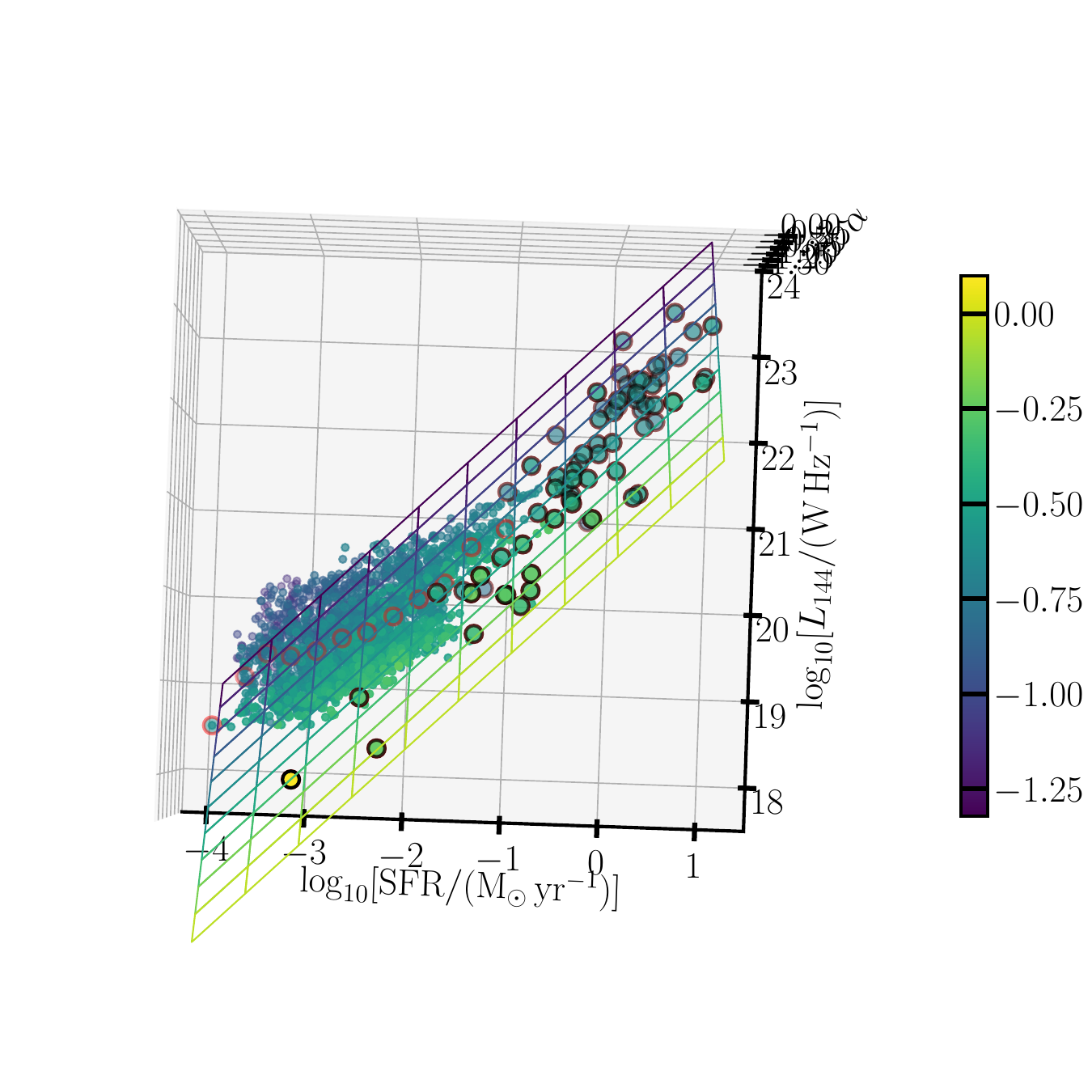}
    \end{subfigure}
    \begin{subfigure}[t]{0.03\linewidth}
        \textbf{(b)}    
    \end{subfigure}
    \begin{subfigure}[t]{0.46\linewidth}
         \includegraphics[width=\linewidth,valign=t]{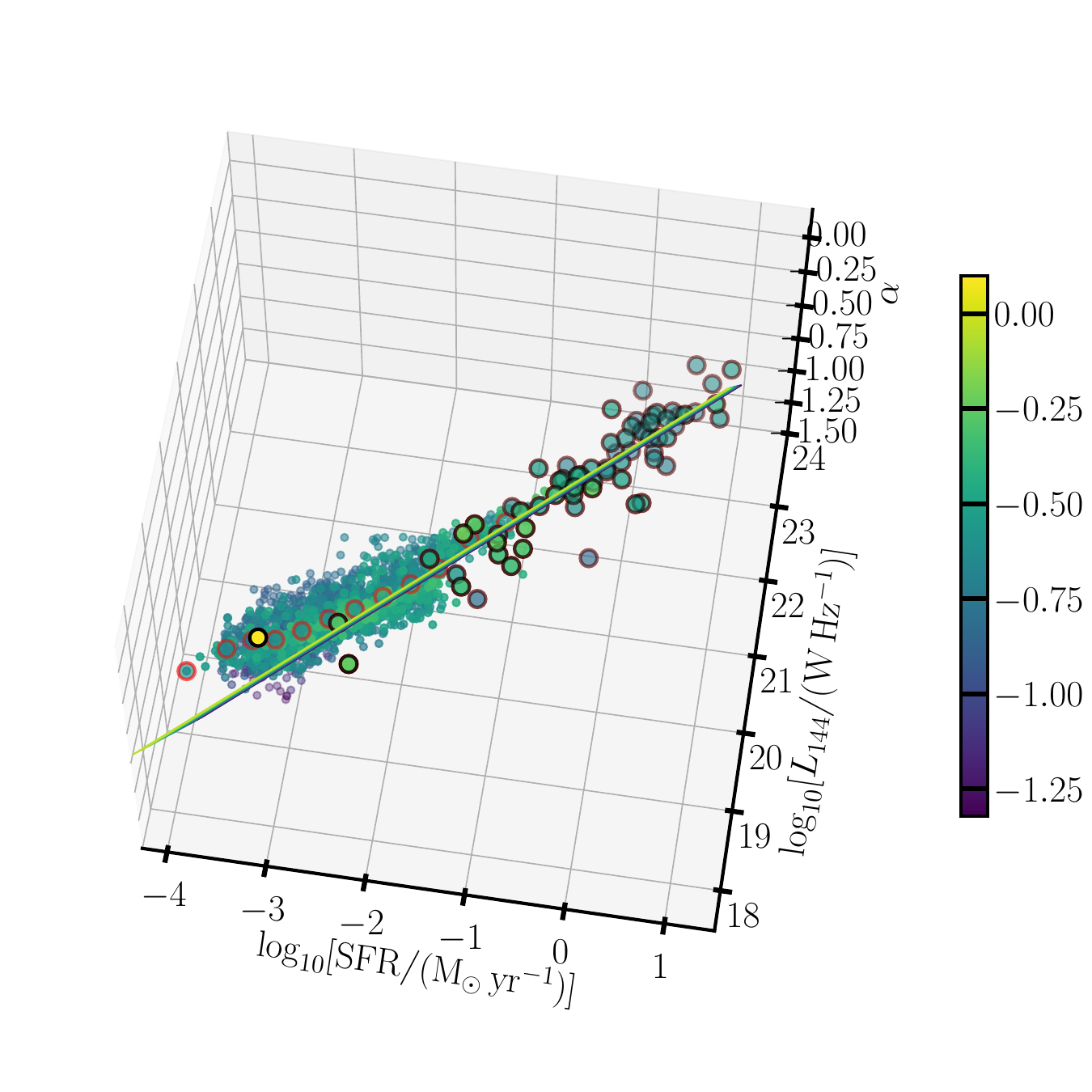}
    \end{subfigure}
    \\
     \begin{subfigure}[t]{0.03\linewidth}
        \textbf{(c)}    
    \end{subfigure}
    \begin{subfigure}[t]{0.46\linewidth}
    \includegraphics[width=\linewidth,valign=t]{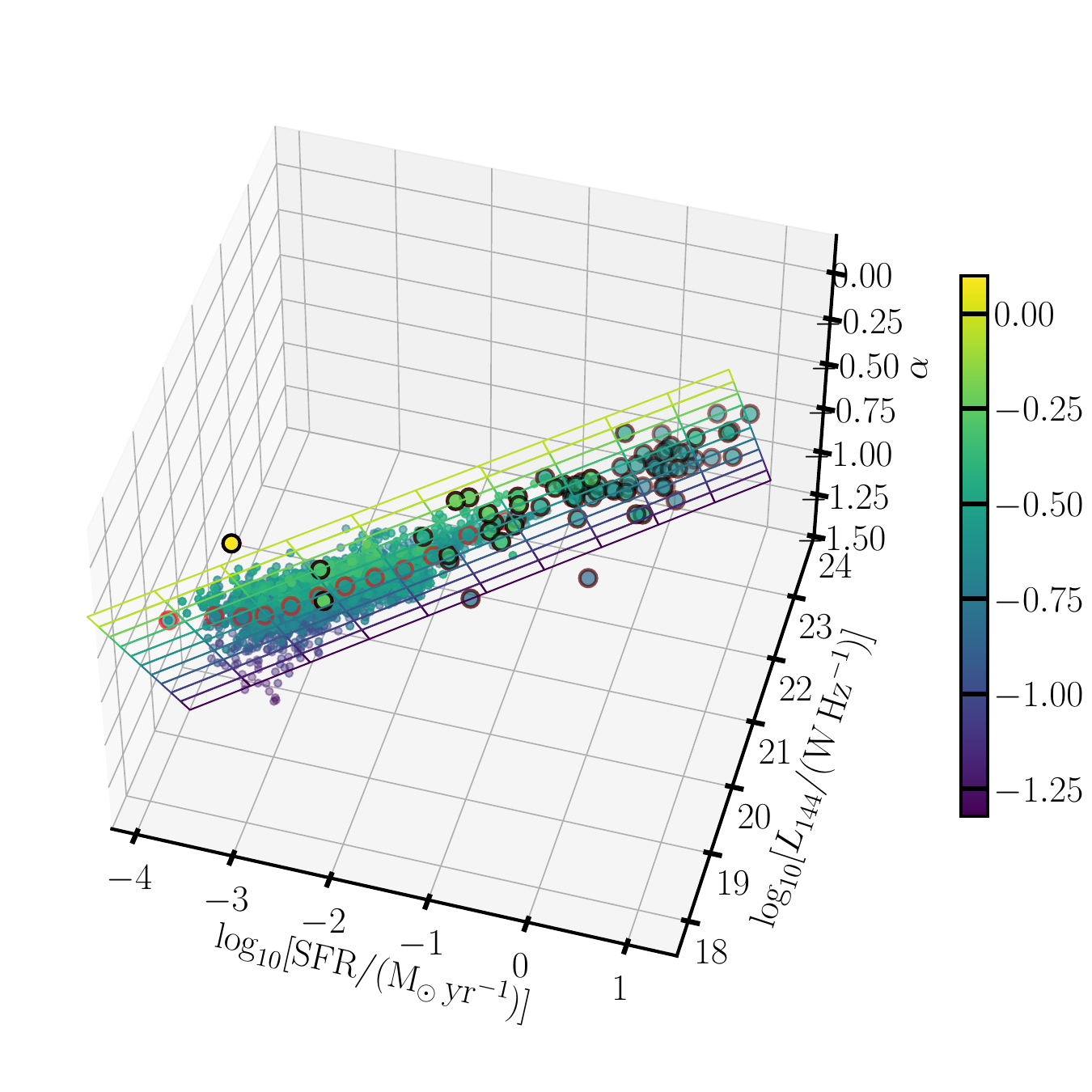}
    \end{subfigure}
    \begin{subfigure}[t]{0.03\linewidth}
        \textbf{(d)}    
    \end{subfigure}
    \begin{subfigure}[t]{0.46\linewidth}
         \includegraphics[width=\linewidth,valign=t]{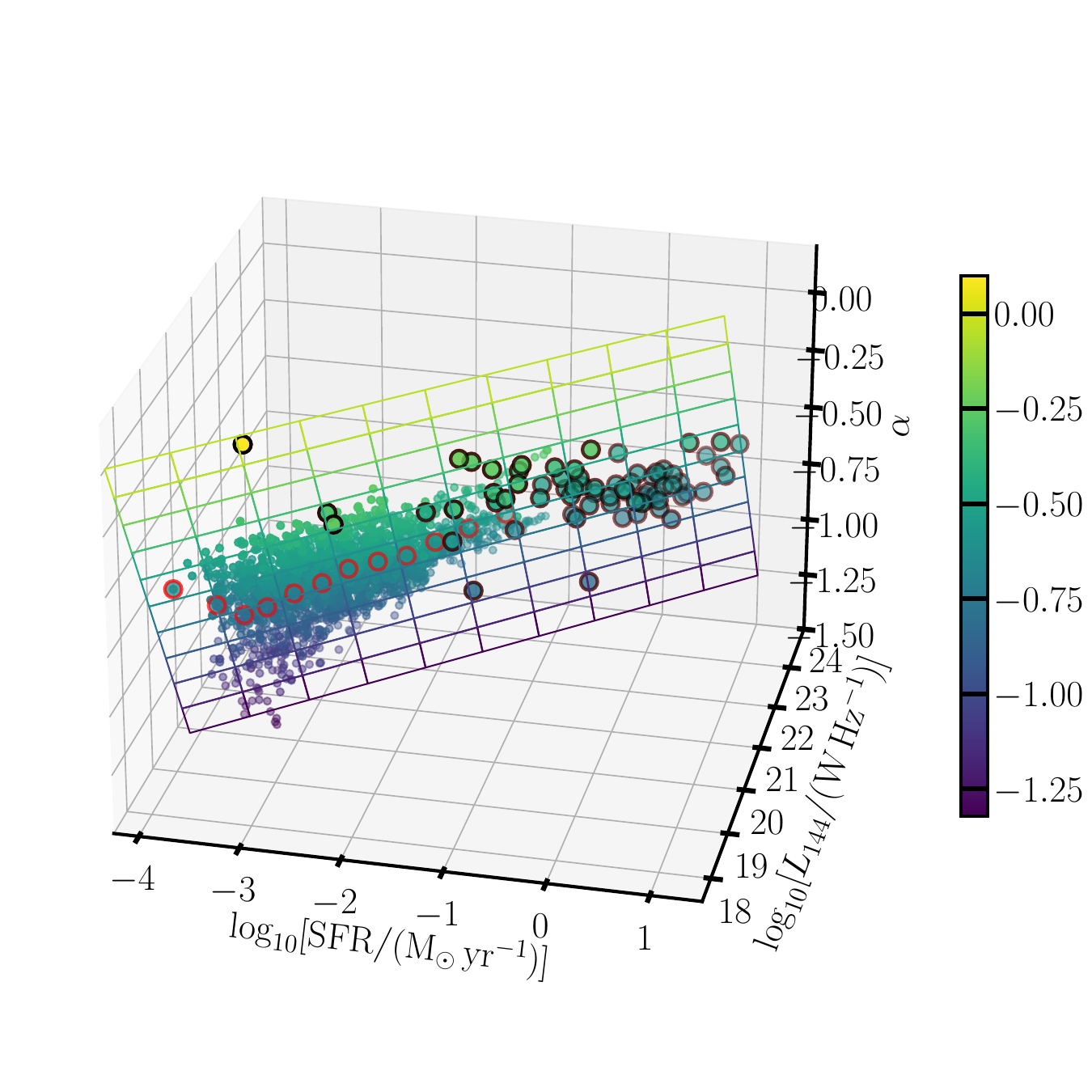}
    \end{subfigure}
    \caption{Fundamental plane of the radio-SFR relation. The grid indicates the position of the fundamental plane; it shows the radio continuum luminosity for an equidistant set of lines in the [log(SFR), $\alpha$] space with the colour coded with radio spectral index.  Large data points with black boundaries show global measurements (individual galaxies) and small data points show local measurements. The colour of the data points is coded with the radio spectral index as indicated by the colour bar.  Panel (a) shows the top view equivalent to the traditional radio-SFR relation (cf.\,Fig.\,\ref{fig:radio_sfr}); panel (b) shows the edge-on representation; panels (c) and (d) show different viewing angles with a perspective from the other side of the plane. An interactive version of this figure can be found \href{https://doi.org/10.5281/zenodo.20395673}{here}.}
    \label{fig:radio_sfr_threed}
\end{figure*}


\subsection{Radio-to-SFR ratio}
\label{ss:radio_to_sfr_ratio}

In this section, we investigate the origin of the radio-to-SFR ratio, which is defined as the ratio of radio continuum luminosity to the SFR. As shown in Sect.\,\ref{ss:radio_sfr_relation}), the radio-to-SFR ratio depends on the radio spectral index. This is shown quite clearly in Fig.\,\ref{fig:ratio_alpha}. We find that galaxies that are fairly radio bright in comparison to their SFR, have a steep radio continuum spectrum. The opposite is true for those galaxies with a flat spectrum. This trend can be observed for the global and the local measurements. The interesting question was whether both measurements follow the same trend. This is approximately the case, although the global measurements have slightly lower ratios than the local measurements at the same radio spectral index. Nevertheless, the agreement is in general quite good.

Hence, the parametrisation of the radio-to-SFR ratio with radio spectral index  using both global and local data is as follows:
\begin{equation}
     \log_{10} (L_{144}) - \log_{10}({\rm SFR}) = (-3.15\pm 0.05)\alpha + (19.94\pm 0.03),
    \label{eq:eta_alpha}
\end{equation}
using again the {\sc odr} method. This result includes both the global and local data. The relation is slightly steeper than the one using only local data \citep{heesen_24a}.

\subsection{Fundamental plane of the radio--SFR relation}

\label{ss:fundamental_plane}

Our results presented in Sect.\,\ref{ss:radio_to_sfr_ratio} show a strong correlation between radio-to-SFR ratio and radio spectral index. In our previous work in \citet{heesen_24a} we had introduced a parametrisation where we correct the radio luminosity using Eq.\,\eqref{eq:eta_alpha}. However, the more general parametrisation is as follows:
\begin{equation}
    \log_{10} (L_{144})  = a \log_{10}({\rm SFR}) + b \alpha + c,
    \label{eq:fundamental_plane}
\end{equation}
where $a$, $b$, and $c$ are constants. The only difference to Eq.\,\eqref{eq:eta_alpha} is that the slope of the radio-SFR relation as function of SFR is now the free parameter $a$; previously we assumed the slope to be unity ($a=1$). We performed a bivariate least-squares fitting of our data with Eq.\,\eqref{eq:fundamental_plane} including both binned local and global measurements. Our best-fitting parameters are $a=1.10\pm 0.06$, $b=-1.75\pm0.29$, and $c=20.90\pm 0.18$. We found a standard deviation (scatter of data points around the best-fitting model) of $0.38$\,dex. This standard deviation is the lowest when compared with the other two-dimensional projections. For $L_{144}$--SFR we found 0.59\,dex; SFR--$\alpha$ has a standard deviation of 1.09\,dex and $L_{144}$--$\alpha$ has one of 1.23\,dex. It is obvious that only a comparison with $L_{144}$--SFR is necessary when considering that it is the only strong correlation with a Spearman's correlation coefficient of $\rho=0.93$ whereas the other correlations have $|\rho|\leq 0.3$.

Hence, we find that the radio luminosity is correlated with both SFR and the radio spectral index $\alpha$, at a highly significant level. In particular, we show for the first time that the data points define a `Fundamental Plane' in the [$\log(L_{144})$, $\log({\rm SFR})$, $\alpha$] space. This is shown in Fig.\,\ref{fig:radio_sfr_threed} where we present a three-dimensional representation of these parameters. We have chosen a view point such that we looking onto the plane at an angle such that the top part is closer to us (with flat radio spectrum in yellow-green colour) than the bottom rear part (with a steep radio spectrum with a dark blue colour).  In general, however, the data points lie in a plane. We note that the two-dimensional projection in the $\log({\rm SFR})$--$\log(L_{144})$ plane is the radio-SFR correlation as presented in Fig\,\ref{fig:radio_sfr}.

There are only a few non-detected galaxies in the sample, but theoretically this could distort the results somewhat since we bias towards galaxies with high radio luminosities at low SFRs. The bias would result in somewhat smaller slopes of the radio-SFR relation as function of SFR. However, because their number (six) is so small, we expect the influence of this bias to be rather small.


\subsection{Unified radio-SFR relation}
\label{ss:unified_radio_sfr_relation}

The three-dimensional representation of the radio-SFR relation in the fundamental plane, allows us to define a unified relation that works both for global and local data. To this end we can choose a best-fit plane on which the most of the data lie. 

\begin{figure}
    \centering
    \includegraphics[width=\linewidth]{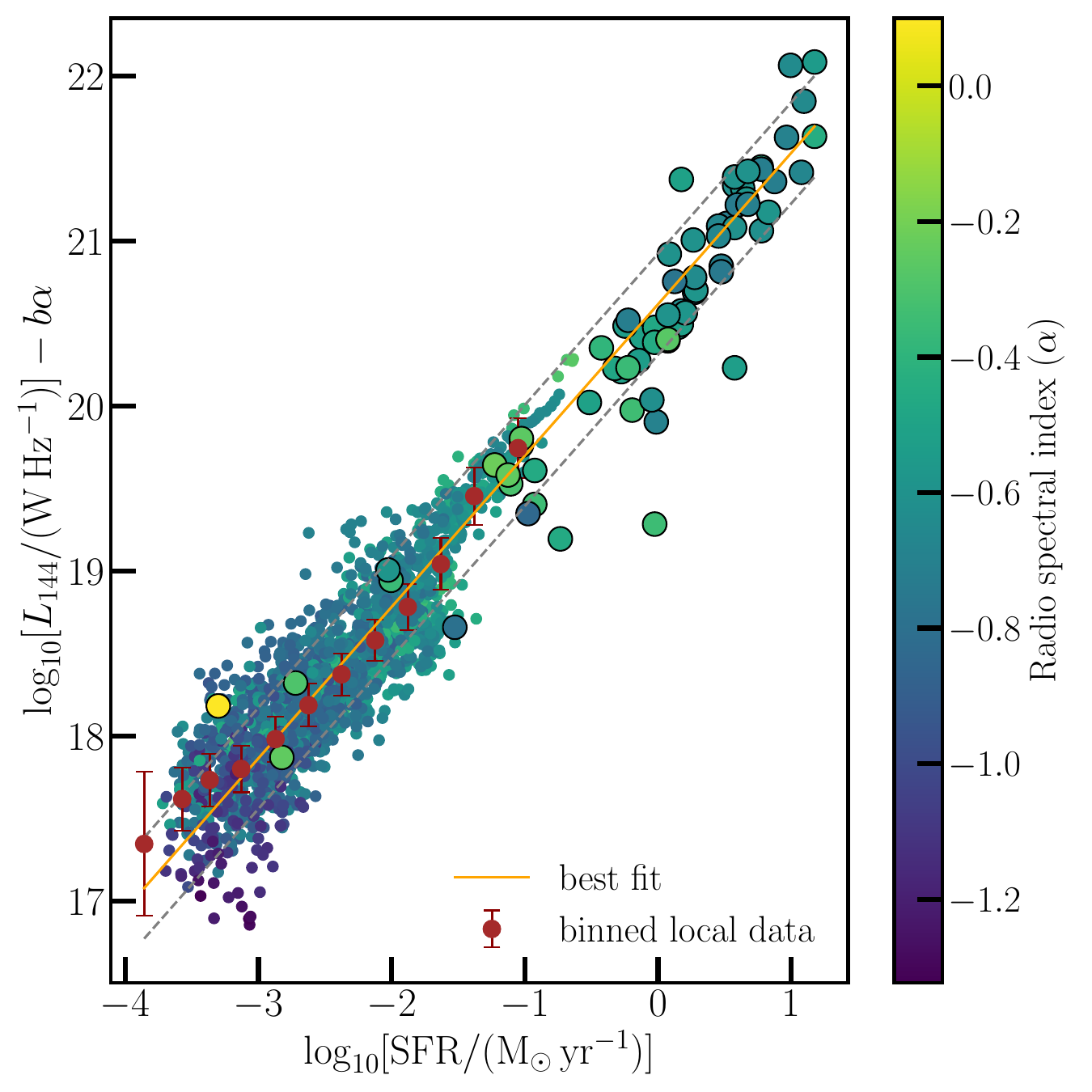}
    \caption{Unified radio-SFR relation. We show the edge-on projection of the fundamental plane of the radio-SFR relation. The $y$-axis shows the radio continuum luminosity corrected for the radio spectral index, $\log(L_{144})-b\alpha$ (cf.\,Eq.\,\ref{eq:fundamental_plane_edge}), and the $x$-axis the SFR. Large data points with black boundaries show global measurements, large data points with red boundaries binned local measurements, and small data points local measurements. The best-fitting relation together with the $\pm 1\sigma_y$ interval are shown as well. To be compared with Fig.\,\ref{fig:radio_sfr}.}
    \label{fig:radio_sfr_unified}
\end{figure}

This is shown in Fig.\,\ref{fig:radio_sfr_unified}. On the $y$-axis, the radio continuum luminosity is divided by a correction factor using the radio spectral index. This can be done by re-writing Eq.\eqref{eq:fundamental_plane}:
\begin{equation}
    \log_{10} (L_{144}) - b \alpha = a \log_{10}({\rm SFR}) + c.
    \label{eq:fundamental_plane_edge}
\end{equation}
Now plotting the left-hand side of Eq.\,\eqref{eq:fundamental_plane_edge} as function of the SFR, we would expect in the ideal case a unified radio-SFR relation where, both, local and global measurements follow the same trend. This is indeed at least approximately the case. The relation is slightly super-linear where global and local measurements now line up quite nicely. Comparing Fig.\,\ref{fig:radio_sfr} with Fig.\,\ref{fig:radio_sfr_unified}, the improvement is remarkable, in particular the agreement of local and global measurements. This means that we can reduce the scatter around the best-fitting relation by a significant amount using two frequencies instead of one, as we effectively do because we are using a two-point radio spectral index. Similar results were already found by \citet{tabatabaei_17a} at gigahertz frequencies for global measurements. The new aspect of our work is that this applies even to local (kiloparsec sized) measurements.

Figure\,\ref{fig:radio_sfr_unified} also shows that the trend of global data is not quite consistent. The global data could be fit with a slightly steeper linear relation than our combined fit. Consequently, the slope $a$ in Eq.\,\eqref{eq:fundamental_plane} depends on the binning of the local data. No binning results into sub-linear slopes. Whereas binning the local data as we did results in slightly super-linear slopes. Hence, we may want to keep in mind that the systematics of the binning may dominate the uncertainties of the best-fitting parameters of Eq.\,\eqref{eq:fundamental_plane}.

\subsection{Influence of fundamental galaxy parameters}
\label{ss:influence_of_fundamental_galaxy_parameters}

\begin{figure}
\includegraphics[width=\linewidth]{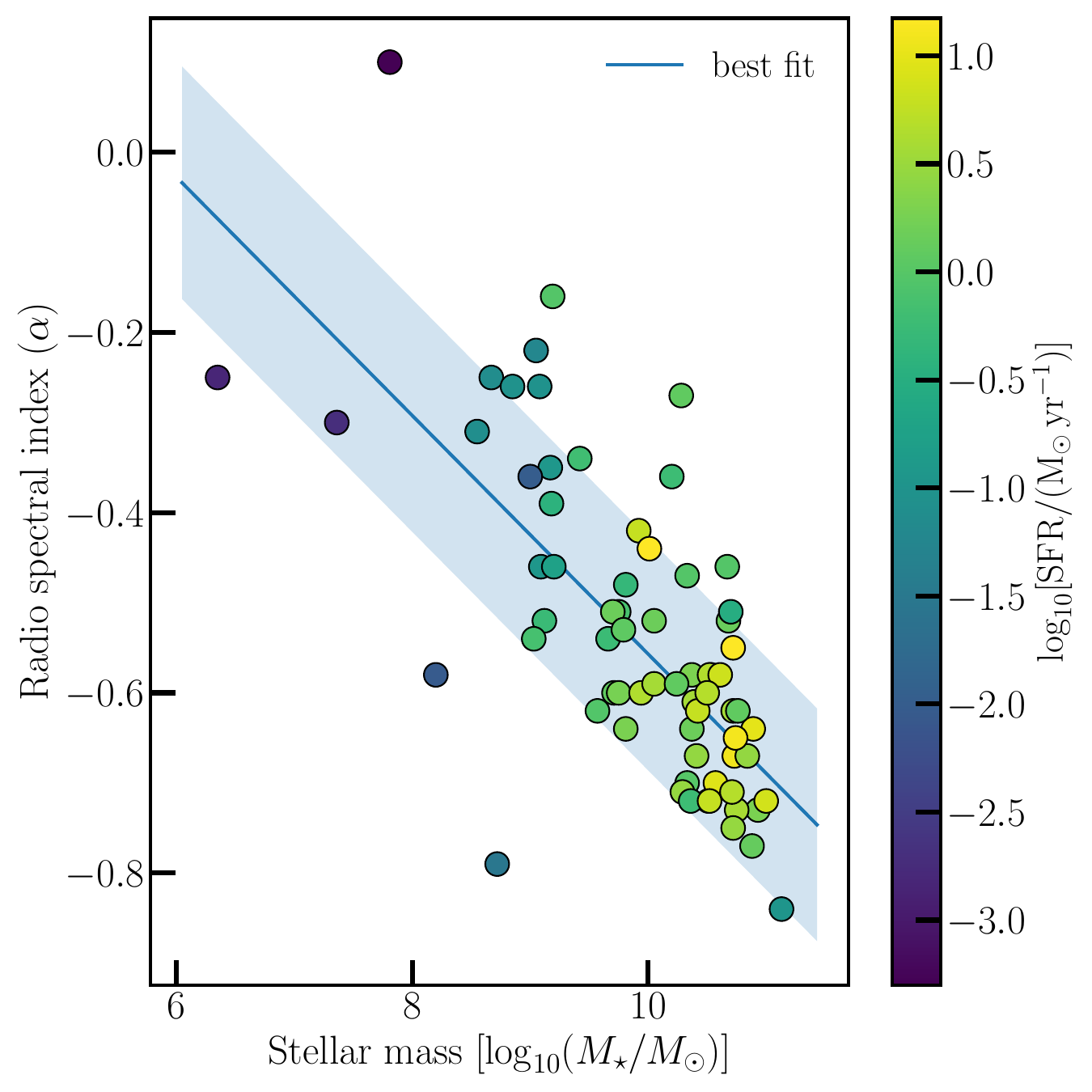}
    \caption{Radio spectral index as a function of stellar mass. each data point is an individual galaxy. The blue line is the best-fitting relation and the shaded areas show $\pm 1\sigma_y$ RMS data scatter intervals. The data points are coloured according to their SFR values.}
    \label{fig:alpha_mass}
\end{figure}

We now investigate the correlations of the global radio spectral index with fundamental galaxy parameters. Best-fitting correlations are presented in Table\,\ref{tab:fit}. We can see clear indications of correlations with SFR, stellar mass, size, and rotation velocity. As an example, we present the correlation with stellar mass in Fig.\,\ref{fig:alpha_mass}. As found in our earlier work \citep{heesen_22a}, the radio continuum spectrum steepens with increasing mass of the galaxy. But, of course, the mass of the galaxy is highly correlated with SFR, size, and  rotation speed. We find that the correlations are equally strong with the exception of the SFR where it is slightly weaker (as indicated by Spearman's rank correlation coefficient). Similar results were found for a different sample of nearby galaxies by \citet{grundy_25a}.

This correlation implies that galaxies with higher masses and higher SFRs have steeper radio continuum spectra. This can explain the super-linear radio-SFR slopes for global measurements. It is unknown which parameter determines the radio spectral index the most but the strongest correlations are with size, mass, and rotation speed. Notably, we find no correlation with the SFR surface density \citep[as in][]{heesen_22a}. In \citet{heesen_22a} we presented a heuristic model that explains why size, and therefore in effect mass, may be the most important parameter while the SFR surface density plays no role. However, \citet{smolinski_26a} find a weak dependence on SFR surface density for the radio spectral index of the thick disc. Also, \citet{tabatabaei_17a} find such a correlation for the global spectral index at gigahertz frequencies. This may imply that other processes play a role at low frequencies, such as free-free absorption \citep{gajovic_25a} and cosmic ray ionisation losses \citep{gajovic_24a} that can flatten the spectrum.

On the other hand no such correlation is found for global measurements between radio spectral index and SFR surface density \citep[][and this work]{heesen_22a}. A correlation is found for gigahertz frequencies \citep{tabatabaei_17a} or for the halo emission at low frequencies \citep{smolinski_26a}.

An important question is whether mass and spectral index are independent parameters for the radio-SFR correlation or whether they are related, and if so, which of the two parameters is more fundamental. We suggest that the spectral index is more fundamental  because the radio power is immediately related to the spectral index, but only indirectly on mass. However, in the future one needs to check whether mass or any other fundamental galaxy parameter also spans another dimension. The assumption that massive galaxies are better calorimeters than low-mass galaxies is based on the assumption that the CR escape rate depends on mass. This suggested by the fact that the radio continuum scale height has a negative correlation with mass surface density \citep{krause_18a,heesen_25a}. However, according to some simulations \citep[e.g.][]{pfrommer_17a} this is controversial and it is unclear whether CRs are passively driven by the baryonic wind, or whether the wind is driven by CRs. In the latter case, no mass dependence of the escape rate is expected and the magnetic field in the halo plays a role.

\section{Discussion}
\label{s:discussion}

\subsection{Cosmic-ray electron calorimetry}
\label{ss:cosmic_ray_electron_calorimetry}

Here, we investigate whether the strong dependence of the radio-to-SFR ratio on the radio spectral index (Sect.\,\ref{ss:radio_to_sfr_ratio}) can be explained within a simplified framework. Conceptually, when cosmic-ray electrons (CREs) escape freely from a galaxy, the observed spectral index equals the injection spectral index, $\alpha_{\rm inj}$. Conversely, if the escape time-scale greatly exceeds the radiative loss time-scale (due to synchrotron and inverse-Compton cooling), the radio continuum spectrum steepens, and the spectral index decreases to $\alpha_{\rm inj}-1$. Hence, for an injection spectral index of $\alpha_{\rm inj}\approx -0.5$ we would expect a radio spectral index between $-1.0$ and $-0.5$.  At the same time, the radio continuum luminosity should decline as CRE escape becomes more efficient. The key question is thus how strongly this effect influences the observed radio emission. 

To explore this, we constructed a simple model of CRE transport in a galactic wind designed to capture the essential physics of this process. Using the SPectral INdex Numerical Analysis of K(c)osmic-ray Electron Radio-emission  \citep[{\sc spinnaker};][]{heesen_16a, heald_22a}, we computed one-dimensional advection models of CRE transport that include only radiative energy losses from synchrotron and inverse-Compton processes.  We assumed a galactic wind speed that is constant with $100$\,\uvel\ and a disc magnetic field strength of $B_0=8\,\muup\rm G$. The halo magnetic field strength is then $B=B_0\exp(-z/h_B)$, where $h_B$ is the magnetic field scale height. Note that we used a fairly generic model that only attempts to reproduce the general features of escaping CRE and neglects the more complicated aspects of galactic winds. However, we do not expect that the details of the wind model to play an important role. Choosing a faster of slower wind speed could be compensated by a larger or smaller magnetic field scale height; accelerating galactic winds \citep[e.g.][]{miskolczi_19a,heald_22a} do also not significantly change this picture, as they only differ in that they include adiabatic losses that  can be described by an energy-independent loss time-scale as the escape; finally, choosing a larger or smaller magnetic field strength can be compensated by a faster or smaller wind speed. Only, when the magnetic field strength would drop below $\approx$$3\,\muup\rm G$ we would expect a difference due to dominating inverse Compton losses in the radiation field of the cosmic microwave background. This would suppress the radio continuum luminosity. We then integrated the radio continuum luminosity along the line-of-sight to obtain the integrated intensity. This intensity is what we see when we look at the galaxy in projection. In the same way we calculated the radio spectral index. For the CRE injection spectral index we choose $p=2.1$ resulting in $\alpha_{\rm inj}=-0.55$. We varied the magnetic field scale height in order to model various degrees of CRE escape.

In Fig.\,\ref{fig:ratio_alpha} we show the resulting radio continuum intensities as a function of radio spectral index. We have varied only the magnetic field scale height using $h_B=2^n\times 0.75\,{\rm kpc}$ with $n=\lbrace 0, 1, \ldots 7 \rbrace$. The model is able to reproduce the general features reasonably well, although there are some limitations at the extreme ends. When the radio spectral index is flat, close to the injection spectral index, the radio continuum intensity drops sharply, consistent with nearly free CRE escape. In contrast, as the radio spectral index approaches $\alpha_{\rm inj}-0.5$, in the limit of strong CRE radiation losses and no escape, the radio continuum luminosity increases. The model successfully explains the observed radio-to-SFR ratio for most spatially resolved data points; however, it underpredicts the ratio for regions with very flat spectra. Similarly, on global scales, galaxies exhibiting unusually flat spectra cannot be reproduced. This limitation arises because the model can only generate spectra approaching the injection slope, but not flatter ones, suggesting that additional physical processes must be at play in such systems. Examples for this are cosmic ray ionisation losses \citep[e.g.][]{gajovic_24a} and free-free absorption \citep[e.g.][]{adebahr_13a, gajovic_25a}. Also, we note that we cannot distinguish between advection and energy-independent diffusion \citep{heesen_23a}.

\subsection{Radio-SFR relation parametrisation}
\label{ss:radio_sfr_relation_parametrisation}

In this section, we present several parametrisations of the radio-SFR relation. In particular, we are using our correlations of the radio spectral index together with the calorimetric efficiency. First we normalise our best-fitting relation to a radio spectral index of $\alpha=-0.55$ which is close to the mean of our global data points which is $\alpha=-0.54\pm 0.17$. This means:

\begin{equation}
    \frac{L_{144}}{\rm W\,Hz^{-1}} = 10^{21.87\pm 0.18 - (1.72\pm 0.29) (\alpha +0.55)} \left (\frac{{\rm SFR}}{\rm M_\sun\,yr^{-1}}\right )^{1.10\pm 0.06}.
    \label{eq:radio_sfr_alpha}
\end{equation}
This normalisation is in good agreement with previous measurements at the same frequency \citep{smith_21a,heesen_22a}. If the spectral index data are available, then Eq.\,\eqref{eq:radio_sfr_alpha} is the preferred way of estimating the SFR. However, since this not generally the case, we present two more useful parametrisations of this general equation below. Entering the dependence of the radio spectral index on mass, we obtain:

\begin{equation}
    \frac{L_{144}}{\rm W\,Hz^{-1}} = 10^{21.87\pm 0.18} \left (\frac{M_{\star}}{10^{10}\rm M_\sun}\right )^{0.23\pm 0.05}   \left (\frac{{\rm SFR}}{\rm M_\sun\,yr^{-1}}\right )^{1.10\pm 0.06}.
    \label{eq:radio_sfr_mass}
\end{equation}
This relation is in good agreement with the previously reported parametrisation of the radio-SFR relation with stellar mass \citep{shenoy_26a}. Finally, we can also express our relation only as function of SFR, when using the correlation between $\alpha$ and SFR:
\begin{equation}
    \frac{L_{144}}{\rm W\,Hz^{-1}} = 10^{21.98\pm 0.18} \left (\frac{{\rm SFR}}{\rm M_\sun\,yr^{-1}}\right )^{1.33\pm 0.08}.
    \label{eq:radio_sfr_sfr}
\end{equation}
This last equation has a normalisation in good agreement with previous work. As expected from earlier works \citep[e.g.][]{heesen_22a}, the relation is significantly super-linear. 


\section{Conclusions}
\label{s:conclusions}

We study nearby galaxies in LoTSS-DR3 \citep{shimwell_26a}, the third data release of the LoTSS \citep{shimwell_17a, shimwell_19a, shimwell_22a}. We included 70 galaxies, significantly expanding our previous sample from LoTSS-DR2 \citep{heesen_22a}. In particular, we included a few dwarf galaxies meaning that we expand the range of SFRs to now five orders of magnitude. We also included in our analysis the spatially resolved data at kiloparsec resolution that we previously analysed in \citet{heesen_24a}. We measured radio continuum luminosities, SFRs, and radio spectral indices. We then investigated the correlations between radio continuum luminosity and SFR, the radio-SFR relation, and their dependence on radio spectral index. Finally, we consider the influence of other fundamental galaxy parameters. These are our main results:

\begin{enumerate}
    \item For the first time, we show that the radio-SFR relation can be best described by a `fundamental plane' which is defined in the ($\log(L)$, $\log(\rm SFR)$, $\alpha$) space. With this description, the distinction between global and local relation vanishes and all data points fall on the same relation. The scatter of the data points around the best-fitting relation is improved from $0.67$\,dex to $0.38$\,dex.
    \item Using an edge-on projection, correcting the radio continuum luminosity for the radio spectral index, we find a slightly super-linear relation with SFR. However, the slope depends on the weighting of global versus local data, with global data having a steeper trend.
    \item The radio-to-SFR ratio, i.e.\ the ratio of radio continuum luminosity to SFR, has a strong dependence on radio spectral index. Galaxies that have steep radio continuum spectrum are brighter in the radio continuum at low frequencies. This can be in part explained by enhanced synchrotron losses of the CRE.
    \item Galaxies with higher masses have steeper radio continuum spectra. With our observed dependence on the radio spectral index, we can therefore explain the reported mass dependence in the literature.
\end{enumerate}

Our results show that it is possible to define a unified radio-SFR relation that hold both for global and local measurements. This unifies the previously reported super-linear relations for global measurements \citep[e.g.][]{heesen_22a} and the sub-linear relations for local measurements \citep[e.g.][]{murphy_08a,heesen_24a}. The latter is ascribed to the influence of cosmic ray transport in a galaxy. Our results show the value of using a two-point radio spectrum rather than at a single frequency. This of course can be extended to even more frequencies \citep{tabatabaei_17a, grundy_25a} in order to gain insights into the emission spectrum. In the future, it would be interesting to study the influence of the varying radio spectrum for instance at higher redshifts in order to better understand star formation rates in the early Universe.
 
Last but not least, great strides have been made to better understand the role of cosmic rays and magnetic fields in galaxy evolution \citep[see e.g.][for recent reviews]{ruszkowski_23a, thompson_24a}. The electron component of the cosmic rays can now be included with injection, transport, and various energy losses. Given the complexity of the physics involved, even the most advanced simulations have to make some simplifying assumptions. Thus, our results can serve as a simple of litmus test to test cosmological simulations with cosmic ray physics.

\begin{acknowledgement}
We thank the anonymous referee for a prompt and constructive report. We thank Rainer Beck and Dan Smith for valuable comments on the manuscript. MB acknowledges funding by the Deutsche Forschungsgemeinschaft (DFG) under Germany's Excellence Strategy -- EXC 2121 ``Quantum Universe" --  390833306 and the DFG Research Group "Relativistic Jets".
MALL  acknowledges grants  PID2024-155875OB-I00 funded by MICIU/AEI/10.13039/501100011033/FEDER, EU  and RYC2020-029354-I funded by MICIU/AEI/10.13039/501100011033 by “ESF Investing in your future”, by “ESF+”. \href{https://www.lofar.org}{LOFAR} is the Low Frequency Array designed and constructed by \href{htts://www.astron.nl}{ASTRON}. It has observing, data processing, and data storage facilities in several countries, which are owned by various parties (each with their own funding sources), and which are collectively operated by the \href{https://www.lofar.eu}{LOFAR ERIC} under a joint scientific policy. The LOFAR resources have benefited from the following recent major funding sources: CNRS-INSU, Observatoire de Paris and Université d'Orléans, France; BMFTR, MKW-NRW, MPG, Germany; Science Foundation Ireland (SFI), Department of Business, Enterprise and Innovation (DBEI), Ireland; NWO, The Netherlands; The Science and Technology Facilities Council, UK; Ministry of Science and Higher Education, Poland; The Istituto Nazionale di Astrofisica (INAF), Italy.

This research made use of the Dutch national e-infrastructure with support of the SURF Cooperative (e-infra 180169) and the LOFAR e-infra group. The Jülich LOFAR Long Term Archive and the German LOFAR network are both coordinated and operated by the Jülich Supercomputing Centre (JSC), and computing resources on the supercomputer JUWELS at JSC were provided by the \href{http://www.gauss-centre.eu}{Gauss Centre for Supercomputing} e.V. (grant CHTB00) through the John von Neumann Institute for Computing (NIC).

This research made use of the \href{https://www.herts.ac.uk/}{University of Hertfordshire high-performance computing facility} and the LOFAR-UK computing facility located at the University of Hertfordshire and supported by STFC [ST/P000096/1], and of the Italian LOFAR IT computing infrastructure supported and operated by INAF, and by the Physics Department of Turin university (under an agreement with Consorzio Interuniversitario per la Fisica Spaziale) at the C3S Supercomputing Centre, Italy.
K.M. acknowledges the support of the Polish National Science Center (NCN) grant UMO-2024/53/B/ST9/00230. 

This research is part of the project LOFAR Data Valorization (LDV) [project numbers 2020.031, 2022.033, and 2024.047] of the research programme Computing Time on National Computer Facilities using SPIDER that is (co-)funded by the Dutch Research Council (NWO), hosted by SURF through the call for proposals of Computing Time on National Computer Facilities.
This research made use of following software packages and other resources: Aladin sky atlas developed at CDS, Strasbourg Observatory, France \citep{bonnarel_00a,boch_14a}; {\sc Astropy} \citep{astropy_13a,astropy_18a};  HyperLeda \citep[\href{http://leda.univ-lyon1.fr}{http://leda.univ-lyon1.fr};][]{makarov_14a}; NASA/IPAC Extragalactic Database (NED), which is operated by the Jet Propulsion Laboratory, California Institute of Technology, under contract with the National Aeronautics and Space Administration; SAOImage DS9 \citep{joye_03a}; and {\sc SciPy} \citep[\href{https://scipy.org}{https://scipy.org};][]{scipy_20a}.

\end{acknowledgement}

\bibliographystyle{aa}
\bibliography{review} 

\appendix

\section{Observational results}
\label{as:results}

In this section, we present observational and derived results for our sample galaxies. These include SFRs and radio continuum luminosities at 144\,MHz, and radio spectral indices between 144 and approximately 1400\,MHz. See Table\,\ref{tab:sample} for the compiled observational results of our sample.

\begin{figure}
    \includegraphics[width=\linewidth,valign=t]{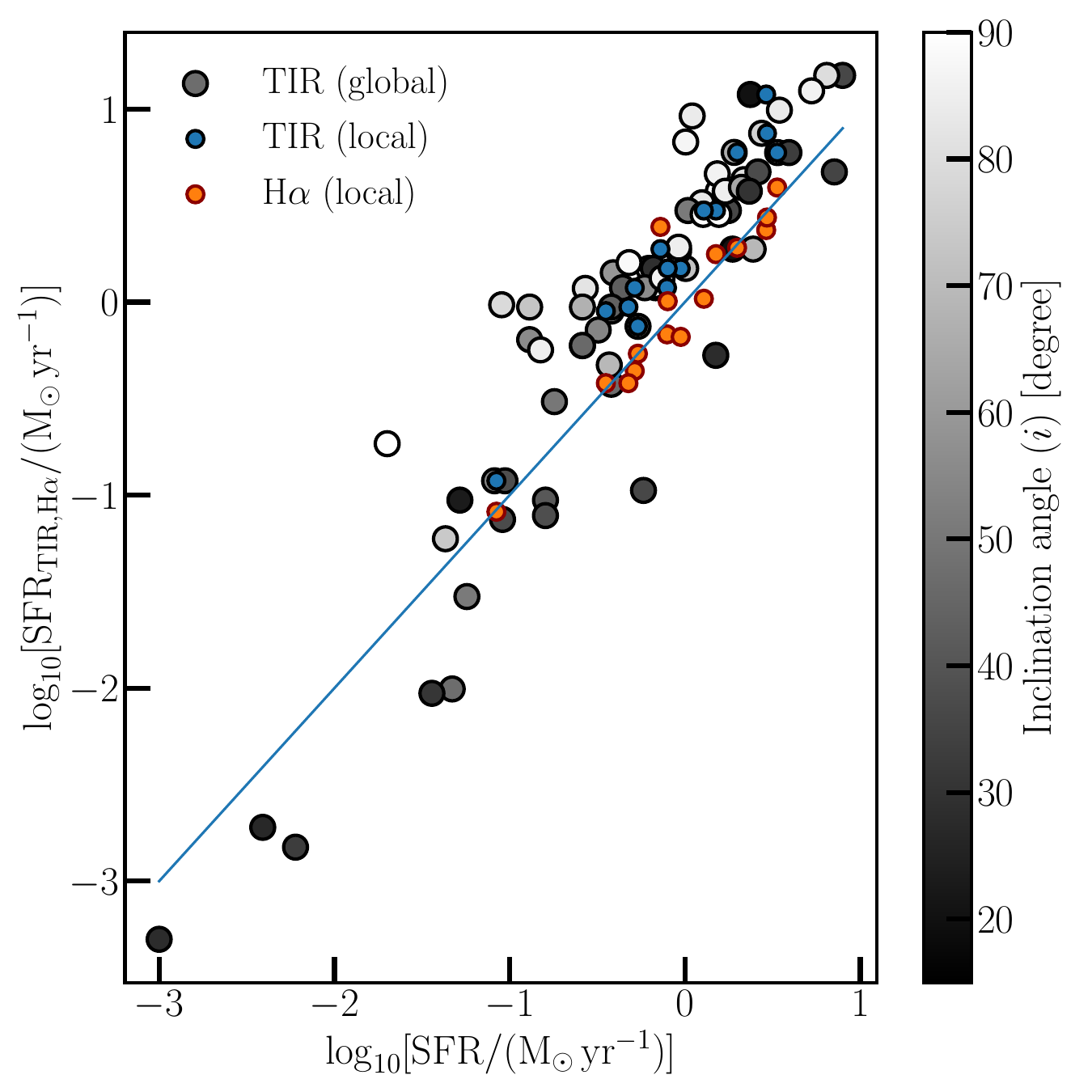}
    \caption{Comparison of SFRs. In this figure we compare the SFRs from the three different methods used. On the $x$-axis we show the SFRs as used in the main part of this work; for global data extinction-corrected H$\alpha$ data and for local data extinction-corrected far-ultraviolet data. On the $y$-axis we show the SFRs for TIR and H$\alpha$. Global data are shown as large grey data points. Local data are shown as small blue (TIR) and orange (H$\alpha$) data points. The 1:1 relation is shown as well as blue line.}
    \label{fig:sfr_compare}
\end{figure}

\subsection{Star formation rates}
\label{as:star_formation_rates}

We calculated SFRs with two different methods. The first method is to use dust-corrected H$\alpha$ luminosities. The SFR is then calculated using the prescription from \citet{kennicutt_12a}. The resulting values are tabulated in \citet{kennicutt_11a}. Again, for those galaxies where the values were not available, we calculated them ourselves. We used H$\alpha$ fluxes and subtracted the contribution from \ion{N}{ii}. Then we used the 24\,$\muup\rm m$ flux density to dust correct the H$\alpha$ luminosity. For NGC\,4125 we did not find a H$\alpha$ flux, so we directly converted the mid-infrared luminosity into the SFR using the conversion from \citet{rieke_09a}. 

Second, we used the total infrared luminosity. The total infrared luminosity refers to a wavelength range of $3$--$1100$\,\micron\ using the values of \citet{kennicutt_11a}, corrected to our distances. These values are based on {\emph Herschel} measurements at 24, 70, and 160\,\micron\ and using the prescription from \citet{dale_09a}. For galaxies that are not part of the KINGFISH sample, we calculated the TIR luminosity using \emph{IRAS} data at 25, 60, and 100\,\micron\ using the prescription from \citet{dale_09a}. The TIR luminosities are then converted into the SFRs using the conversion from \citet{kennicutt_12a}, which is based on both \citet{hao_11a} and \citet{murphy_11a}.

Detailed values with references used for calculating the SFRs for both methods can be found in Table\,\ref{tab:sfr1}.


\subsection{Results with TIR SFRs}
\label{as:tir_results}

\begin{figure}[!t]
    \includegraphics[width=\linewidth]{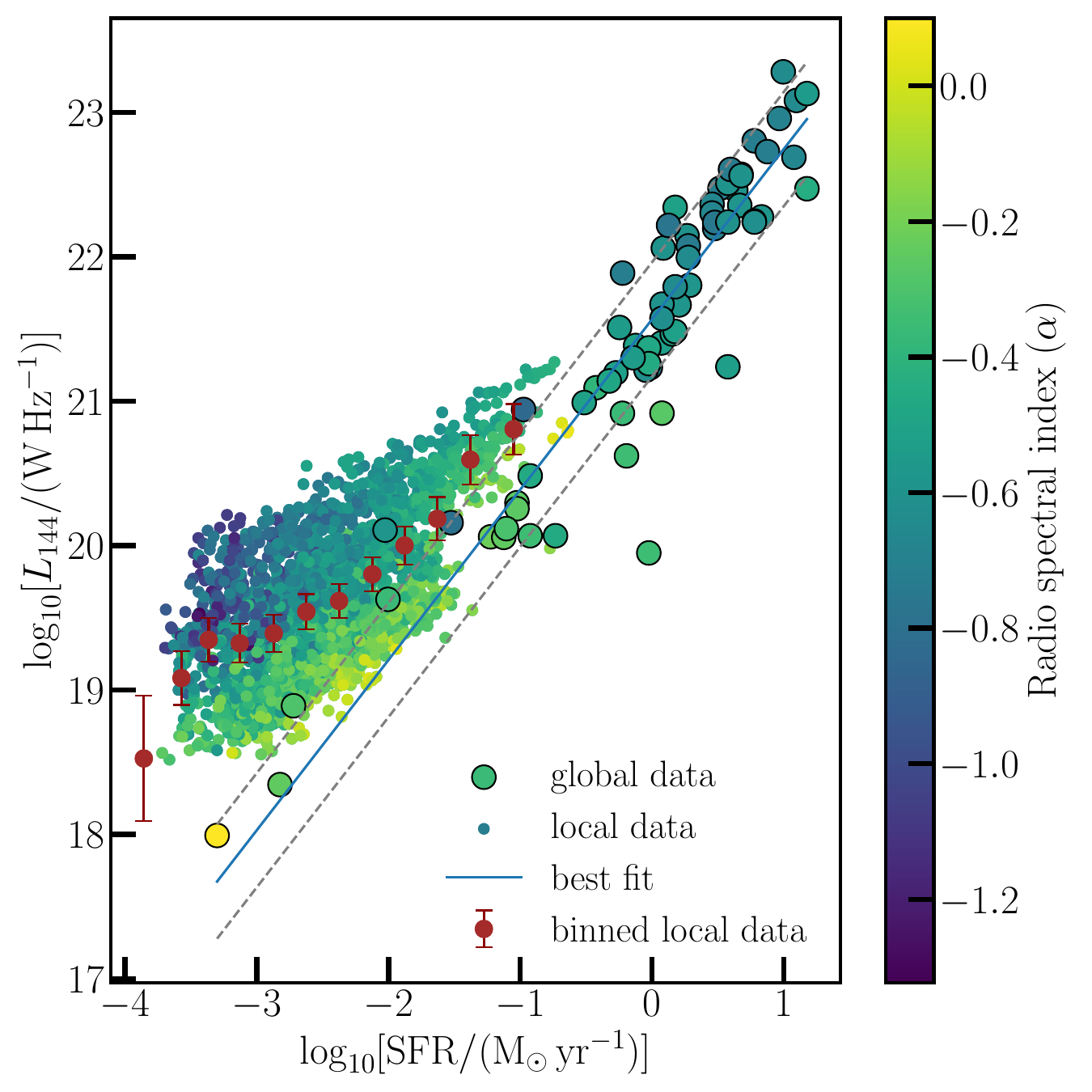}
    \caption{Radio-SFR relation. Same as Fig.\,\ref{fig:radio_sfr} but using TIR-based SFR values.}
    \label{fig:radio_sfr_tir}
\end{figure}

\begin{figure}[!t]
    \includegraphics[width=\linewidth]{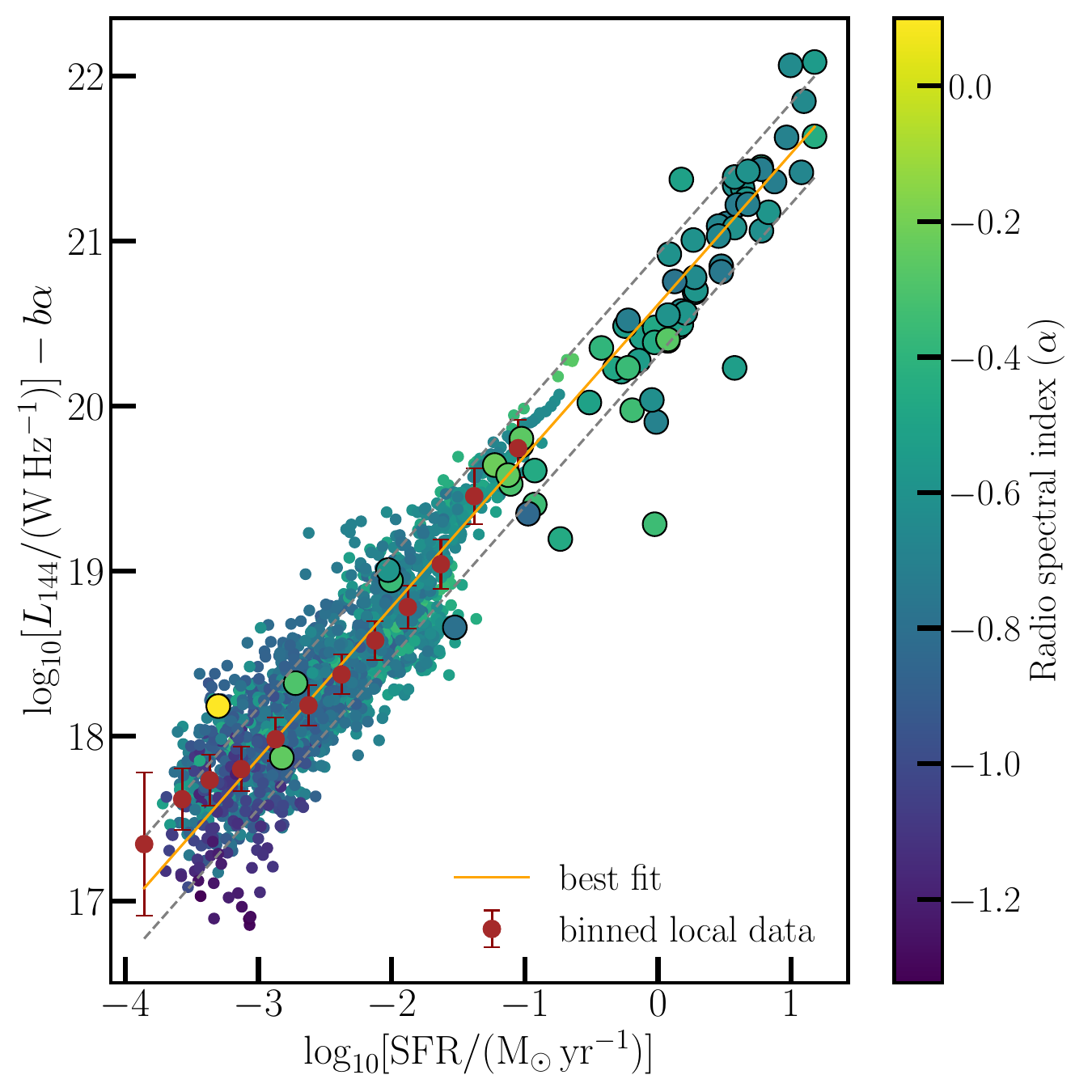}
    \caption{Unified radio-SFR relation. Same as Fig.\,\ref{fig:radio_sfr_unified} but using TIR-based SFR values.}
    \label{fig:radio_sfr_unified_tir}
\end{figure}

When using the TIR SFR values for the global data points, we obtain a smaller dependency on the SFR. When fitting Eq.\,\eqref{eq:radio_sfr} we obtain a slope of only $1.18\pm 0.05$. The resulting figure is presented in Fig.\,\ref{fig:radio_sfr_tir}. The smaller power-law slope can be understood as a consequence of lower-mass systems. In them, the TIR luminosity underestimates the true SFR as they are low in dust content. Similarly, when fitting the fundamental plane (Eq.\,\ref{eq:fundamental_plane}) we obtain $a=0.92\pm 0.04$, $b=-1.72\pm 0.25$, and $c=20.62\pm 0.14$. Again, the dependence on SFR is much smaller, in fact now even sub-linear. The corresponding unified radio-SFR relation is shown in Fig.\,\ref{fig:radio_sfr_unified_tir}. 

Finally, the ratio radio-to-SFR relation as function of radio spectral index becomes:
\begin{equation}
     \log_{10} (L_{144}) - \log_{10}({\rm SFR}) = (-3.18\pm 0.05)\alpha - (19.91\pm 0.03),
    \label{eq:eta_alpha_tir}
\end{equation}
This relation is very similar to the results using the H$\alpha$ and mid-infrared based SFRs. The reason is that the fit is dominated by the local data points that are identical.

\subsection{Comparison of SFRs}
\label{as:comparison_of_sfrs}

For our work we ideally would like to choose SFRs that are consistent for the entire sample, regardless whether this applies to global or local data. In principle, mid-infrared data is not a good star formation tracer in edge-on galaxies since the radiation becomes optically thick \citep[e.g.][]{vargas_19a}. Therefore, H$\alpha$ corrected luminosities underestimate the SFRs in highly inclined galaxies. This is shown in Fig.\,\ref{fig:sfr_compare}, where one sees that highly inclined galaxies have larger TIR SFRs than obtained from H$\alpha$. On the other hand, at the lowest SFRs, the TIR-based SFRs are too low. The reason is that dwarf galaxies are low in dust and hence, some dust-heating photons escape the galaxy. Nevertheless, since we have many highly inclined galaxies in the sample from the CHANGE-ES survey, using would make sense from that perspective.

Now we look at the local SFRs. We now compare the sum of the local SFRs with the global SFRs in those 15 galaxies where we have local data. In the ideal case, they should agree. We do find, however, that the local SFRs are a factor of a few lower than the global SFRs in the same galaxies, if we use TIR-based global SFRs. This is shown as small blue data points in Fig.\,\ref{fig:sfr_compare}. It is presently unclear why this deviation exists as these are not highly inclined galaxies. On the contrary, if we use H$\alpha$-based SFRs instead, we find a good agreement between local and global SFRs. This is shown as orange data points in Fig.\,\ref{fig:sfr_compare}. Hence, on this basis we have chosen to use the SFRs from H$\alpha$ for the global data points. In the next section (Sect.\,\ref{as:tir_results}), present alternative results when using TIR-based SFRs.

\subsection{Radio spectral indices}
\label{as:radio_spectral_indices}

Radio spectral indices were calculated between 144 and approximately 1400\,MHz. For this we used mostly published flux densities from the literature. See Table\,\ref{tab:spix1} for details. In a few cases re-measured the flux densities from maps at the second frequency. For this we used the same integration ellipse as for the LOFAR data. The major and minor semi-axes of these ellipses together with the position angle of the major axes can be found in Table\,\ref{tab:sample}. We also used the projected semi-major axis as the distance of the galaxy to measure the star-forming radius $r_\star$. With this the SFR surface density is calculated as $\Sigma_{\rm SFR}={\rm SFR}/(\piup r_\star^2)$. Rotation velocities with references can be found in \citet{heesen_22a}.

\begin{table*}
\centering
\caption{Properties of galaxies in the sample.}
\label{tab:sample}
\begin{tabular}{l cccccccc c}
\hline\hline
Galaxy & $d$ & $\log_{10}(\rm SFR)$ & $\log_{10}(\rm SFR_{\rm TIR})$ & $\log_{10}(L_{144})$ & $\alpha$ & $a_{20}$ & $b_{20}$ & $pa$ & $\log_{10} (M_{\star})$ \\
& (Mpc) & (\usfr) & (\usfr) & ($\rm W\,Hz^{-1}$) & & ($\arcmin$) & ($\arcmin$) & ($\degr$) & $(\rm M_\sun)$ \\  
(1) & (2)  & (3) & (4) & (5) & (6) &  (7) & (8) & (9) & (10) \\ \hline
IC\,10 & $0.7$ & $-1.33$ & $-2.00$ & $19.63\pm0.04$ & $-0.36\pm0.04$ & $5.0$ & $4.0$ & $129$ & $9.00$ \\ 
NGC\,598 & $0.9$ & $-0.89$ & $-0.19$ & $20.62\pm0.04$ & $-0.34\pm0.05$ & $20.0$ & $13.0$ & $23$ & $9.42$ \\ 
NGC\,628 & $7.2$ & $-0.17$ & $0.07$ & $21.67\pm0.04$ & $-0.59\pm0.05$ & $5.2$ & $4.9$ & $20$ & $10.24$ \\ 
NGC\,855 & $9.73$ & $-1.37$ & $-1.23$ & $20.06\pm0.04$ & $-0.22\pm0.09$ & $0.7$ & $0.6$ & $67$ & $9.05$ \\ 
NGC\,891 & $9.1$ & $0.19$ & $0.57$ & $22.51\pm0.04$ & $-0.62\pm0.04$ & $5.9$ & $4.6$ & $22$ & $10.72$ \\ 
NGC\,925 & $9.12$ & $-0.27$ & $-0.13$ & $21.39\pm0.04$ & $-0.51\pm0.06$ & $5.6$ & $3.1$ & $287$ & $9.75$ \\ 
IC\,342 & $3.28$ & $0.27$ & $0.27$ & $22.00\pm0.04$ & $-0.64\pm0.09$ & $17.0$ & $12.0$ & $37$ & $9.81$ \\ 
NGC\,2146 & $17.2$ & $0.90$ & $1.17$ & $23.13\pm0.04$ & $-0.55\pm0.04$ & $2.6$ & $2.4$ & $123$ & $10.72$ \\ 
NGC\,2403 & $3.06$ & $-0.42$ & $-0.05$ & $21.21\pm0.04$ & $-0.62\pm0.06$ & $11.1$ & $5.1$ & $124$ & $9.57$ \\ 
Ho\,II & $3.05$ & $-1.44$ & $-2.03$ & $20.11\pm0.04$ & $-0.58\pm0.05$ & $3.0$ & $2.5$ & $15$ & $8.20$ \\ 
DDO\,53 & $3.61$ & $-2.22$ & $-2.82$ & $18.34\pm0.04$ & $-0.25\pm0.06$ & $0.9$ & $0.5$ & $114$ & $6.35$ \\ 
NGC\,2683 & $6.27$ & $-1.05$ & $-0.01$ & $21.23\pm0.04$ & $-0.70\pm0.06$ & $3.6$ & $2.4$ & $44$ & $10.33$ \\ 
NGC\,2798 & $25.8$ & $0.53$ & $0.77$ & $22.25\pm0.04$ & $-0.42\pm0.06$ & $1.1$ & $0.8$ & $160$ & $9.92$ \\ 
NGC\,2820 & $26.5$ & $-0.21$ & $0.17$ & $22.34\pm0.04$ & $-0.51\pm0.04$ & $3.2$ & $1.2$ & $65$ & $9.70$ \\ 
NGC\,2841 & $14.1$ & $0.39$ & $0.27$ & $22.08\pm0.04$ & $-0.73\pm0.05$ & $3.9$ & $2.9$ & $153$ & $10.93$ \\ 
NGC\,2903 & $8.9$ & $0.28$ & $0.78$ & $22.24\pm0.04$ & $-0.62\pm0.05$ & $5.1$ & $4.1$ & $204$ & $10.42$ \\ 
Ho\,I & $3.9$ & $-2.41$ & $-2.72$ & $18.89\pm0.04$ & $-0.30\pm0.10$ & $2.0$ & $1.0$ & $50$ & $7.36$ \\ 
NGC\,2976 & $3.55$ & $-1.09$ & $-0.93$ & $20.48\pm0.04$ & $-0.46\pm0.05$ & $3.0$ & $2.8$ & $143$ & $9.09$ \\ 
NGC\,3003 & $25.4$ & $-0.17$ & $0.08$ & $22.06\pm0.04$ & $-0.60\pm0.06$ & $3.1$ & $1.5$ & $79$ & $9.71$ \\ 
NGC\,3031 & $3.44$ & $-0.41$ & $0.15$ & $21.47\pm0.04$ & $-0.52\pm0.05$ & $11.7$ & $5.9$ & $330$ & $10.68$ \\ 
NGC\,3034 & $3.52$ & $0.81$ & $1.17$ & $22.47\pm0.04$ & $-0.44\pm0.05$ & $6.2$ & $7.0$ & $66$ & $10.01$ \\ 
NGC\,3079 & $20.6$ & $0.54$ & $0.99$ & $23.28\pm0.04$ & $-0.64\pm0.04$ & $3.9$ & $2.3$ & $167$ & $10.89$ \\ 
NGC\,3077 & $3.83$ & $-1.03$ & $-0.93$ & $20.07\pm0.04$ & $-0.35\pm0.06$ & $1.4$ & $1.2$ & $49$ & $9.17$ \\ 
M81\,DwB & $3.6$ & $-3.00$ & $-3.30$ & $17.99\pm0.04$ & $0.10\pm0.06$ & $0.6$ & $0.3$ & $139$ & $7.81$ \\ 
NGC\,3184 & $11.7$ & $-0.18$ & $0.17$ & $21.79\pm0.04$ & $-0.64\pm0.05$ & $4.1$ & $4.1$ & $179$ & $10.37$ \\ 
NGC\,3198 & $14.1$ & $0.00$ & $0.17$ & $21.48\pm0.04$ & $-0.52\pm0.07$ & $3.4$ & $1.5$ & $215$ & $10.05$ \\ 
IC\,2574 & $3.79$ & $-1.24$ & $-1.53$ & $20.16\pm0.04$ & $-0.79\pm0.08$ & $6.0$ & $3.6$ & $56$ & $8.72$ \\ 
NGC\,3265 & $19.6$ & $-0.42$ & $-0.43$ & $21.09\pm0.04$ & $-0.39\pm0.06$ & $0.6$ & $0.7$ & $73$ & $9.18$ \\ 
Mrk\,33 & $15.4$ & $0.18$ & $-0.28$ & $21.20\pm0.04$ & $-0.52\pm0.07$ & $0.8$ & $0.7$ & $129$ & $9.12$ \\ 
NGC\,3351 & $9.33$ & $-0.23$ & $0.07$ & $20.92\pm0.04$ & $-0.27\pm0.11$ & $0.7$ & $0.4$ & $11$ & $10.28$ \\ 
NGC\,3432 & $9.42$ & $-0.82$ & $-0.25$ & $21.51\pm0.04$ & $-0.54\pm0.04$ & $3.4$ & $1.9$ & $33$ & $9.66$ \\ 
NGC\,3448 & $24.5$ & $-0.04$ & $0.26$ & $22.15\pm0.04$ & $-0.60\pm0.04$ & $1.8$ & $1.2$ & $65$ & $9.75$ \\ 
NGC\,3556 & $14.09$ & $0.34$ & $0.63$ & $22.46\pm0.04$ & $-0.60\pm0.04$ & $4.4$ & $3.2$ & $79$ & $9.94$ \\ 
NGC\,3627 & $9.38$ & $0.23$ & $0.57$ & $21.24\pm0.04$ & $-0.53\pm0.05$ & $4.2$ & $3.1$ & $173$ & $10.67$ \\ 
NGC\,3628 & $8.5$ & $0.00$ & $0.83$ & $22.28\pm0.04$ & $-0.58\pm0.05$ & $6.9$ & $3.4$ & $104$ & $10.61$ \\ 
NGC\,3735 & $42.0$ & $0.04$ & $0.96$ & $22.96\pm0.04$ & $-0.70\pm0.04$ & $2.3$ & $1.5$ & $130$ & $10.57$ \\ 
NGC\,3773 & $12.4$ & $-0.80$ & $-1.03$ & $20.30\pm0.04$ & $-0.26\pm0.06$ & $0.6$ & $0.6$ & $161$ & $8.85$ \\ 
NGC\,3877 & $17.7$ & $-0.04$ & $0.28$ & $21.80\pm0.04$ & $-0.58\pm0.04$ & $2.1$ & $1.0$ & $35$ & $10.37$ \\ 
NGC\,3938 & $17.9$ & $0.25$ & $0.47$ & $22.20\pm0.04$ & $-0.71\pm0.05$ & $3.0$ & $3.0$ & $0$ & $10.29$ \\ 
NGC\,4013 & $16.0$ & $-0.32$ & $0.20$ & $21.67\pm0.04$ & $-0.58\pm0.04$ & $2.1$ & $0.9$ & $65$ & $10.53$ \\ 
NGC\,4096 & $10.32$ & $-0.57$ & $0.07$ & $21.40\pm0.04$ & $-0.53\pm0.04$ & $3.2$ & $1.6$ & $20$ & $9.79$ \\ 
NGC\,4125 & $24.77$ & $-0.24$ & $-0.98$ & $20.94\pm0.04$ & $-0.84\pm0.06$ & $0.3$ & $0.3$ & $95$ & $11.13$ \\ 
NGC\,4157 & $15.6$ & $0.10$ & $0.51$ & $22.47\pm0.04$ & $-0.72\pm0.04$ & $4.8$ & $2.7$ & $66$ & $10.51$ \\ 
NGC\,4214 & $2.95$ & $-0.80$ & $-1.11$ & $20.12\pm0.04$ & $-0.31\pm0.06$ & $2.7$ & $1.4$ & $320$ & $8.55$ \\ 
NGC\,4217 & $20.6$ & $0.18$ & $0.66$ & $22.36\pm0.04$ & $-0.58\pm0.04$ & $2.4$ & $1.8$ & $50$ & $10.52$ \\ 
NGC\,4236 & $4.45$ & $-0.89$ & $-0.03$ & $19.95\pm0.04$ & $-0.35\pm0.05$ & $5.6$ & $2.4$ & $161$ & $9.19$ \\ 
NGC\,4244 & $4.4$ & $-1.70$ & $-0.73$ & $20.07\pm0.04$ & $-0.46\pm0.04$ & $6.2$ & $1.1$ & $45$ & $9.20$ \\ 
NGC\,4254 & $14.4$ & $0.59$ & $0.77$ & $22.80\pm0.09$ & $-0.72\pm0.05$ & $4.5$ & $4.0$ & $56$ & $10.52$ \\ 
NGC\,4321 & $14.3$ & $0.42$ & $0.67$ & $22.57\pm0.09$ & $-0.71\pm0.05$ & $3.5$ & $3.2$ & $153$ & $10.71$ \\ 
NGC\,4449 & $4.02$ & $-0.49$ & $-0.14$ & $21.30\pm0.04$ & $-0.54\pm0.06$ & $4.6$ & $4.6$ & $51$ & $9.03$ \\ 
NGC\,4450 & $14.07$ & $-0.74$ & $-0.52$ & $20.99\pm0.09$ & $-0.51\pm0.08$ & $0.8$ & $0.6$ & $155$ & $10.70$ \\ 
NGC\,4559 & $6.98$ & $-0.43$ & $-0.33$ & $21.14\pm0.04$ & $-0.48\pm0.07$ & $4.7$ & $2.2$ & $148$ & $9.81$ \\ 
NGC\,4565 & $11.9$ & $-0.13$ & $0.12$ & $22.22\pm0.04$ & $-0.77\pm0.04$ & $8.0$ & $2.5$ & $135$ & $10.88$ \\ 
NGC\,4625 & $9.3$ & $-1.28$ & $-1.03$ & $20.26\pm0.04$ & $-0.26\pm0.15$ & $0.9$ & $0.7$ & $330$ & $9.08$ \\ 
NGC\,4631 & $7.62$ & $0.23$ & $0.57$ & $22.51\pm0.04$ & $-0.59\pm0.04$ & $8.8$ & $7.5$ & $86$ & $10.05$ \\ 
NGC\,4725 & $11.9$ & $-0.36$ & $0.07$ & $21.57\pm0.04$ & $-0.62\pm0.06$ & $3.8$ & $2.4$ & $36$ & $10.76$ \\ 
NGC\,4736 & $4.66$ & $-0.42$ & $-0.03$ & $21.37\pm0.04$ & $-0.47\pm0.05$ & $3.6$ & $2.9$ & $105$ & $10.33$ \\ 
NGC\,4826 & $5.27$ & $-0.59$ & $-0.23$ & $20.92\pm0.04$ & $-0.36\pm0.06$ & $1.8$ & $1.1$ & $121$ & $10.20$ \\ 
NGC\,5033 & $17.13$ & $0.03$ & $0.59$ & $22.61\pm0.04$ & $-0.73\pm0.05$ & $5.0$ & $3.4$ & $172$ & $10.75$ \\ 
\hline
\end{tabular}
\end{table*}

\setcounter{table}{0}
\begin{table*}
\centering
\caption{Properties of galaxies in the sample.}
\label{tab:sample2}
\begin{tabular}{l cccccccc c}
\hline\hline
Galaxy & $d$ & $\log_{10}(\rm SFR)$ & $\log_{10}(\rm SFR_{\rm TIR})$ & $\log_{10}(L_{144})$ & $\alpha$ & $a_{20}$ & $b_{20}$ & $pa$ & $\log_{10} (M_{\star})$ \\
& (Mpc) & (\usfr) & (\usfr) & ($\rm W\,Hz^{-1}$) & & ($\arcmin$) & ($\arcmin$) & ($\degr$) & $(\rm M_\sun)$ \\  
(1) & (2)  & (3) & (4) & (5) & (6) &  (7) & (8) & (9) & (10) \\ \hline
NGC\,5055 & $7.94$ & $0.02$ & $0.47$ & $22.24\pm0.04$ & $-0.75\pm0.05$ & $6.6$ & $4.9$ & $102$ & $10.72$ \\ 
NGC\,5194 & $8.0$ & $0.37$ & $1.08$ & $22.69\pm0.04$ & $-0.67\pm0.04$ & $6.9$ & $6.7$ & $195$ & $10.73$ \\ 
NGC\,5195 & $8.0$ & $-0.59$ & $-0.22$ & $21.89\pm0.04$ & $-0.72\pm0.06$ & $2.0$ & $1.8$ & $79$ & $10.36$ \\ 
NGC\,5297 & $40.4$ & $0.10$ & $0.45$ & $22.36\pm0.04$ & $-0.67\pm0.04$ & $2.5$ & $1.2$ & $147$ & $10.41$ \\ 
NGC\,5457 & $6.7$ & $0.37$ & $0.57$ & $22.24\pm0.04$ & $-0.61\pm0.06$ & $11.3$ & $11.3$ & $39$ & $10.39$ \\ 
NGC\,5474 & $6.8$ & $-1.04$ & $-1.13$ & $20.06\pm0.04$ & $-0.25\pm0.06$ & $2.2$ & $1.6$ & $100$ & $8.67$ \\ 
NGC\,5775 & $28.9$ & $0.72$ & $1.09$ & $23.08\pm0.04$ & $-0.65\pm0.04$ & $2.9$ & $1.8$ & $148$ & $10.74$ \\ 
NGC\,5866 & $15.3$ & $-0.59$ & $-0.03$ & $21.26\pm0.04$ & $-0.46\pm0.06$ & $1.0$ & $0.6$ & $127$ & $10.67$ \\ 
NGC\,5907 & $16.8$ & $0.19$ & $0.45$ & $22.30\pm0.04$ & $-0.67\pm0.04$ & $5.8$ & $2.0$ & $156$ & $10.84$ \\ 
NGC\,6946 & $6.8$ & $0.85$ & $0.67$ & $22.56\pm0.04$ & $-0.60\pm0.04$ & $7.3$ & $6.4$ & $243$ & $10.50$ \\ 
NGC\,7331 & $14.5$ & $0.44$ & $0.87$ & $22.73\pm0.04$ & $-0.72\pm0.05$ & $5.4$ & $2.8$ & $168$ & $11.00$ \\ 
\hline
\end{tabular}
\tablefoot{Column (2) $d$ is the distance \citep[see][for references]{heesen_22a}; (3) ; SFR from H$\alpha$ and 24\,\micron\ mid-infrared data (Sect.\,\ref{as:star_formation_rates}); (4) SFR estimated from the TIR luminosity (Sect.\,\ref{as:star_formation_rates}); (5) radio continuum luminosity at 144\,MHz; (6) radio spectral index between 144 and approximately 1400\,MHz (Sect.\,\ref{as:radio_spectral_indices}); (7) semi-major axis of flux integration ellipe; (8) semi-minor axis of flux integration ellipse; (9) position angle of the semi-major axis of the integration ellipse (counting from North over East to South); (10) stellar mass \citep{leroy_19a}. \\  {\bf References for position angles:} Hyperleda, except NGC\,925, 2403, 2841, 2903, 3031, 3184, 3198, 3627, 4826, 5055, 6946, 7731, and IC\,2574 \citep{de_blok_08a}, NGC\,628 \citep{kamphuis_92a}, IC\,342 \citep{crosthwaite_00a}, NGC\,3079, 3877 , and 4157 \citep{krause_18a}, NGC\,4214 and 4625 \citep{lee_11a}, NGC\,5194 \citep{heesen_21a}, NGC\, 5457 \citep{bosma_81a}, NGC\,3938 (assumed to be zero), NGC\,4254 and 4321 (this work).}
\end{table*}

\begin{table*}
\centering
\caption{Total infrared and dust-corrected H$\alpha$ luminosities for the galaxies in the sample.}
\label{tab:sfr1}
\begin{tabular}{l ccccc ccc ccc}
\hline\hline
Galaxy	 & $S_{25}$ & $S_{60}$ & $S_{100}$ & $\log_{10}(L_{\rm TIR})$ & Ref. &  $\log_{10}(F_{\rm H\alpha+\ion{N}{ii}})$ & $\rm \ion{N}{ii}/H\alpha$ & Ref.  & $S_{24}$ &  Ref. & $\log_{10}(\rm L_{H\alpha, \rm corr})$  \\
& (Jy)   & (Jy)    & (Jy)    & $(L_\sun)$&       &  $(\rm erg\,s^{-1}\, cm^{-2})$ &       &        &        (Jy) &       & $(L_\sun)$ \\
(1)      & (2)  & (3)   & (4)   & (5)   &  (6)    &  (7)     & (8)     &  (9)   & (10)    & (11) & (12) \\\hline 
IC 10	 & 4.00	& 31.00	& 71.00 & 7.82  & I86     &  $-9.91$ & 0       &	      &	9.81	&  B12 & 6.36 \\
NGC 598	 &    	&     	&      	& 9.63  & D05	  &          &         &	      &	50   	&  E05 & 6.80 \\
NGC 628	 &    	&     	&      	& 9.90  & K11	  &          &         &	      &		&  K11 & 7.52 \\	        	       
NGC 855	 &    	&     	&      	& 8.60  & K11	  & $-12.23$ & 0.185   &	K09   &	0.082	&  D05 & 6.32 \\
NGC 891	 & 7.00	& 66.46	& 172.23& 10.40 & S03     &          & 	       &	      &	     	&  V19 & 7.96 \\	        
NGC 925	 &  	&     	&      	& 9.70  & K11     & $-11.1$  & 0.201   &	K09   &	0.9  	&  D05 & 7.42 \\
IC 342	 &    	&     	&      	& 10.10 & K11	  &          &         &	      &	   	&  K11 & 7.96 \\	        	       
NGC 2146 &    	&     	&      	& 11.00 & K11	  &          &         &	      &	   	&  K11 & 8.59 \\	        	       
NGC 2403 &    	&     	&      	&  9.78 & D09     & $-10.25$ & 0.088   &	K09   &	5.64 	&  D05 & 7.27 \\
Ho II	 &    	&     	&      	&  7.80 & K11	  &          &         &	      &	   	&  K11 & 6.24 \\	        	       
DDO 53	 &    	&     	&      	&  7.00 & K11	  &          &         &	      &	   	&  K11 & 5.47 \\	        	       
NGC 2683 &    	&     	&      	&  9.81 & D09	  &          &         &	      &	   	&  V19 & 7.08 \\	        	       
NGC 2798 &    	&     	&      	& 10.60 & K11	  &          &         &	      &	   	&  K11 & 8.22 \\	        	       
NGC 2820 & 	& 3.00  & 10.00 & 10.00 & M89	  &          &         &	      &	   	&  V19 & 7.82 \\	           	
NGC 2841 &    	&     	&      	& 10.10 & K11     &          &         &	      &	     	&  K11 & 8.08 \\	           
NGC 2903 &    	&     	&      	& 10.60 & D09     &          &         &	      &	     	&  C10 & 7.97 \\	           
Ho I	 &    	&     	&      	&  7.10 & K11     & $-12.44$ & 0.075   &	K09   &	0.013	&  D05 & 5.28 \\
NGC 2976 &    	&     	&      	&  8.90 & K11     & $-11.19$ & 0.357   &	K09   &	1.33 	&  D05 & 6.60 \\
NGC 3003 &    	& 3.00 	& 9.00  &  9.91 & M86	  &          &         &	      &	   	&  V19 & 7.88 \\	           	
NGC 3031 &    	&     	&      	&  9.98 & D09     & $-10.32$ & 0.545   &	K09   &	4.94 	&  D05 & 7.28 \\
NGC 3034 &    	&     	&      	& 11.00 & D09     &          &         &	      &	     	&  C10 & 8.50 \\	           
NGC 3079 &    	&     	&      	& 10.82 & E08     &          &         &	      &	     	&  V19 & 8.39 \\	           
NGC 3077 &    	&     	&      	& 8.90  & K11     &          &         &	      &	     	&  K11 & 6.66 \\	           
M81 DwB	 &    	&     	&      	& 6.50  & K11     &          &         &	      &	     	&  K11 & 4.69 \\	           
NGC 3184 &    	&     	&      	& 10.00 & K11     &          &         &	      &	     	&  K11 & 7.51 \\	           
NGC 3198 &    	&     	&      	& 10.00 & K11     &          &         &	      &	     	&  K11 & 7.69 \\	           
IC 2574	 &    	&     	&      	& 8.30  & K11     & $-11.23$ & 0.046   &	K09   &	0.27 	&  D05 & 6.44 \\
NGC 3265 &    	&     	&      	& 9.40  & K11     &          &         &	      &	     	&  K11 & 7.27 \\	           
Mrk 33	 & 1.00	& 5.00 	& 5.00  & 9.55  & M89	  &          &         &	      &	   	&  K11 & 7.86 \\	           	
NGC 3351 &    	&     	&      	& 9.90  & K11     & $-11.24$ & 0.655   &	K09   &	2.4  	&  D05 & 7.46 \\
NGC 3432 &    	&     	&      	& 9.58  & D05     &          &         &	      &	     	&  V19 & 7.39 \\	           
NGC 3448 & 0.64	& 6.64 	& 11.17 & 10.09 & S03     &          & 	       & 	      &	     	&  V19 & 7.94 \\	        
NGC 3556 & 4.19	& 32.55	& 76.90 & 10.46 & S03     &          & 	       &	      &	     	&  V19 & 8.24 \\	        
NGC 3627 &    	&     	&      	& 10.40 & K11     & $-10.74$ & 0.55    &	K09   &	7.25 	&  D05 & 7.92 \\
NGC 3628 &    	&     	&      	& 10.66 & D05     &          &         &	      &	     	&  V19 & 7.84 \\	           
NGC 3735 & 1.12	& 6.71 	& 17.83 & 10.79 & S03     &          & 	       &	      &	     	&  V19 & 8.48 \\	        
NGC 3773 &    	&     	&      	&  8.80 & K11     & $-11.99$ & 0.233   &	K09   &	0.13 	&  D05 & 6.89 \\
NGC 3877 & 1.10	& 7.72 	& 22.42 & 10.11 & S03     &          & 	       &	      &	     	&  V19 & 7.82 \\	        
NGC 3938 &    	&     	&      	& 10.30 & K11     &          &	       &	      &	     	&  K11 & 7.94 \\	           
NGC 4013 & 0.77	& 7.01 	& 24.36 & 10.03 & S03     &          & 	       &	      &	     	&  V19 & 7.54 \\	        
NGC 4096 &    	&     	&      	&  9.90 &  D05    &          &         &	      &	   	&  C10 & 7.54 \\		           
NGC 4125 &	& 1.00 	& 1.00  & 8.85  & I86     &          &         &	      &	0.069	&  D05 & 7.45 \\	
NGC 4157 & 2.12	& 17.71	& 50.67 & 10.34 & S03     &          & 	       &	      &	     	&  V19 & 7.93 \\	        
NGC 4214 & 2.46	& 17.57	& 29.08 &  8.72 & S03     &          &         &	      &	     	&  H18 & 6.89 \\	        
NGC 4217 & 1.50	& 11.60	& 41.19 & 10.49 & S03     &          & 	       &	      &	     	&  V19 & 7.96 \\	        
NGC 4236 &    	&     	&      	&  9.80 & K11     &          &         &	      &	     	&  K11 & 6.80 \\	           
NGC 4244 &    	&     	&      	&  9.09 & D05     &          &         &	      &	     	&  V19 & 6.47 \\	           
NGC 4254 &    	&     	&      	& 10.60 & K11     & $-10.89$ & 0.42    &	K09   &	4.09 	&  D05 & 8.28 \\
NGC 4321 &    	&     	&      	& 10.50 & K11     & $-11.06$ & 0.43    &	K09   &	3.33 	&  D05 & 8.10 \\    
NGC 4449 &    	&     	&      	&  9.68 & D05     & $-12.21$ & 0.51    &	K09   &	0.19 	&  D05 & 7.19 \\
NGC 4450 &    	& 1.00 	& 7.00  &  9.31 & I86     &          &         &	      &	     	&  C10 & 6.94 \\        
NGC 4559 &    	&     	&      	&  9.50 & K11  	  &          &	       &	      &	     	&  K11 & 7.26 \\       
NGC 4565 & 1.36	& 7.79 	& 34.62 &  9.95 & S03	  &          &	       &	      &	     	&  V19 & 7.67 \\        
NGC 4625 &    	&     	&      	&  8.80 & K11  	  &          &	       &	      &	     	&  K11 & 6.40 \\           
NGC 4631 &    	&     	&      	&  10.40& K11  	  &          &	       &	      &	     	&  K11 & 7.92 \\           
NGC 4725 &    	&     	&      	&  9.90 & K11  	  &          &	       &	      &	     	&  K11 & 7.33 \\           
NGC 4736 &    	&     	&      	&  9.80 & K11  	  &          &	       &	      &	     	&  K11 & 7.27 \\           
NGC 4826 &    	&     	&      	&  9.60	& K11     & $-11.15$ & 0.72    &	K09   &	2.47 	&  D05 & 7.10 \\   
NGC 5033 & 2.14	& 16.20	& 50.23 & 10.42 & S03  	  & $-11.23$ & 0.48    &	K09   &	1.92 	&  D05 & 7.71 \\   
\hline                                                                                                   
\end{tabular}                                                                                            
\end{table*}

\setcounter{table}{1}
\begin{table*}
\centering
\caption{Total infrared and dust-corrected H$\alpha$ luminosities for the galaxies in the sample.}
\label{tab:sfr2}
\begin{tabular}{l ccccc ccc ccc}
\hline\hline
Galaxy	 & $S_{25}$ & $S_{60}$ & $S_{100}$ & $\log_{10}(L_{\rm TIR})$ & Ref. &  $\log_{10}(F_{\rm H\alpha+\ion{N}{ii}})$ & $\rm \ion{N}{ii}/H\alpha$ & Ref.  & $S_{24}$ &  Ref. & $\log_{10}(\rm L_{H\alpha, \rm corr})$  \\
& (Jy)   & (Jy)    & (Jy)    & $(L_\sun)$&       &  $(\rm erg\,s^{-1}\, cm^{-2})$ &       &        &        (Jy) &       & $(L_\sun)$ \\
(1)      & (2)  & (3)   & (4)   & (5)   &  (6)    &  (7)   & (8)     &  (9)   & (10)    & (11) & (12) \\\hline 
NGC 5055 &    	&     	&      	& 10.30 & K11	  & $-10.8$  & 0.486 &	K09   &	5.59 	&  D05 & 7.70 \\
NGC 5194 &    	&     	&      	& 10.90 & D05	  & $-10.45$ & 0.59  &	K09   &	12.25	&  D05 & 8.06 \\
NGC 5195 &    	&     	&      	&  9.60 & D05	  &          &       &	      &	1.31   	&  D05 & 7.10 \\	
NGC 5297 &    	& 2.00 	& 8.00  & 10.28 & I86     &          &       &	      &	     	&  V19 & 8.16 \\	        
NGC 5457 &    	&     	&      	& 10.40 & K11	  &          &       &	      &	   	&  K11 & 8.05 \\	        	       
NGC 5474 &    	&     	&      	&  8.70 & K11	  & $-11.65$ & 0.22  &	K09   &	0.18 	&  D05 & 6.65 \\
NGC 5775 & 2.47	& 23.59	& 55.64 & 10.92 & S03	  &          &	     &	      &		&  V19 & 8.57 \\
NGC 5866 &    	&     	&      	&  9.80 & K11	  &          &	     &	      &		&  K11 & 7.10 \\
NGC 5907 & 1.44	& 9.14 	& 37.46 & 10.28 & S03	  &          &	     &	      &		&  V19 & 8.03 \\
NGC 6946 &    	&     	&      	& 10.50 & D05	  &          &	     &	      &		&  K11 & 8.54 \\
NGC 7331 &    	&     	&      	& 10.70 & D05	  &          &	     &	      &		&  K11 & 8.12 \\
\hline
\end{tabular}
\tablefoot{Column (2)--(4) infrared flux density at 25, 60, and 100\,$\muup\rm m$; (5) total infrared luminosity at 2-1000\,$\muup\rm m$; (7) H$\alpha$ flux in $\rm erg\,s^{-1}\,cm^{-2}$; (8) fraction of NII; (10) flux density at 24\,$\muup\rm m$; (12) dust corrected H$\alpha$ flux\\
References: B12: \citet{bendo_12a}; C10: \citet{calzetti_10a}; D05: \citet{dale_05a}; E05: \citet{engelbracht_05a}; H19: \citet{hindson_18a}; I86: IPAC 1986;  K09: \citet{kennicutt_09a}; K11: \citet{kennicutt_11a}; S03: \citet{sanders_03a}; V19: \citet{vargas_19a}.}
\end{table*}

\begin{table*}
\centering
\caption{Radio spectral indices between 144 and approximately 1400\,MHz.}
\label{tab:spix1}
\begin{tabular}{l ccccc}
\hline\hline
Galaxy  & $S_{144}$ & $\nu$ & $S_\nu$  &  $\alpha$   &  Ref. \\
& (Jy)               & (MHz) & (Jy) & & \\\hline
IC\,10	  & $0.721 \pm  0.072$  & 1580 & $0.3060\pm  0.0100$ & $-0.36\pm   0.04$ & H18    \\
NGC\,598  & $4.304\pm  0.431$  &1400  & $1.9907\pm  0.0995$ & $-0.34\pm   0.05$ &   T07  \\
NGC\,628  & $0.756 \pm  0.076$  & 1365 & $0.2000\pm  0.0100$ & $-0.59\pm   0.05$ & B07     \\  
NGC\,855  & $0.010\pm  0.001$  &1400  & $0.0060\pm  0.0010$ & $-0.22\pm   0.09$ &   NVSS \\  
NGC\,891  & $3.294\pm  0.329$  &1575  & $0.7432\pm  0.0105$ & $-0.62\pm   0.04$ &   W15  \\
NGC\,925  & $0.243\pm  0.024$  &1365  & $0.0780\pm  0.0080$ & $-0.51\pm   0.06$ &   B07  \\ 
IC\,342	  & $7.668 \pm  0.767$  & 1400 & $1.8000\pm  0.3000$ & $-0.64\pm   0.09$ & T17    \\
NGC\,2146 & $3.813 \pm  0.381$  & 1365 & $1.1000\pm  0.0100$ & $-0.55\pm   0.04$ & B07     \\ 
NGC\,2403 & $1.457 \pm  0.146$  & 1365 & $0.3600\pm  0.0300$ & $-0.62\pm   0.06$ & B07     \\
Ho\,II    & $0.115 \pm  0.011$  & 1365 & $0.0313\pm  0.0016$ & $-0.58\pm   0.05$ & B07     \\   
DDO\,53	 & $0.001 \pm  0.000$  & 1400 & $0.0008\pm  0.0001$ & $-0.25\pm   0.06$ & B17   \\
NGC\,2683 & $0.362\pm  0.036$  &1575  & $0.0680\pm  0.0070$ & $-0.70\pm   0.06$ &   W15  \\
NGC\,2798 & $0.222\pm  0.022$  &1400  & $0.0860\pm  0.0080$ & $-0.42\pm   0.06$ &   Y01  \\   
NGC\,2820 & $0.261\pm  0.026$  &1575  & $0.0764\pm  0.0012$ & $-0.51\pm   0.04$ &   W15  \\
NGC\,2841 & $0.501\pm  0.050$  &1365  & $0.0970\pm  0.0070$ & $-0.73\pm   0.05$ &   B07  \\
NGC\,2903	 & $1.835 \pm  0.183$  & 1365 & $0.4600\pm  0.0100$ & $-0.62\pm   0.05$ & B07     \\
Ho\,I	 & $0.004 \pm  0.000$  & 4800 & $0.0015\pm  0.0005$ & $-0.30\pm   0.10$ & T17    \\
NGC\,2976 & $0.201\pm  0.020$  &1365  & $0.0720\pm  0.0050$ & $-0.46\pm   0.05$ &   B07  \\
NGC\,3003 & $0.148\pm  0.015$  &1575  & $0.0349\pm  0.0041$ & $-0.60\pm   0.06$ &   W15  \\
NGC\,3031 & $2.055\pm  0.206$  &1400  & $0.6240\pm  0.0312$ & $-0.52\pm   0.05$ &   W92 \\  
NGC\,3034	 & $19.914\pm  1.991$  & 1365 & $7.4100\pm  0.3700$ & $-0.44\pm   0.05$ & A13    \\  
NGC\,3079 & $3.746\pm  0.375$  &1575  & $0.8075\pm  0.0162$ & $-0.64\pm   0.04$ &   W15  \\
NGC\,3077 & $0.067\pm  0.007$  &1400  & $0.0300\pm  0.0030$ & $-0.35\pm   0.06$ &   Y01  \\  
M81DwB	 & $0.001 \pm  0.000$  & 1400 & $0.0008\pm  0.0001$ & $0.10 \pm   0.06$ & B17   \\  
NGC\,3184 & $0.377\pm  0.038$  &1365  & $0.0890\pm  0.0050$ & $-0.64\pm   0.05$ &   B07  \\
NGC\,3198 & $0.127\pm  0.013$  &1365  & $0.0390\pm  0.0050$ & $-0.52\pm   0.07$ &   B07  \\
IC\,2574	 & $0.083 \pm  0.008$  & 1365 & $0.0152\pm  0.0020$ & $-0.76\pm   0.07$ & B17     \\
NGC\,3265 & $0.027\pm  0.003$  &1400  & $0.0110\pm  0.0010$ & $-0.39\pm   0.06$ &   Y01  \\  
Mrk\,33   & $0.055\pm  0.006$  &1400  & $0.0170\pm  0.0020$ & $-0.52\pm   0.07$ &   Y01  \\  
NGC3351	 & $0.079 \pm  0.008$  & 1400 & $0.0430\pm  0.0100$ & $-0.27\pm   0.11$ & T17    \\  
NGC\,3432 & $0.304\pm  0.030$  &1575  & $0.0833\pm  0.0019$ & $-0.54\pm   0.04$ &   W15  \\
NGC\,3448 & $0.195\pm  0.019$  &1575  & $0.0465\pm  0.0009$ & $-0.60\pm   0.04$ &   W15  \\
NGC\,3556 & $1.215\pm  0.122$  &1575  & $0.2909\pm  0.0058$ & $-0.60\pm   0.04$ &   W15  \\
NGC\,3627	 & $1.636 \pm  0.164$  & 1365 & $0.5000\pm  0.0100$ & $-0.53\pm   0.05$ & B07     \\  
NGC\,3628	 & $2.177 \pm  0.218$  & 1365 & $0.5900\pm  0.0100$ & $-0.58\pm   0.05$ & B07     \\
NGC\,3735	 & $0.429 \pm  0.043$  & 1575 & $0.0813\pm  0.0016$ & $-0.70\pm   0.04$ & W15   \\
NGC\,3773	 & $0.011 \pm  0.001$  & 1400 & $0.0060\pm  0.0006$ & $-0.26\pm   0.06$ & B17   \\
NGC\,3877 & $0.169\pm  0.017$  &1575  & $0.0423\pm  0.0009$ & $-0.58\pm   0.04$ &   W15  \\
NGC\,3938 & $0.409\pm  0.041$  &1365  & $0.0820\pm  0.0050$ & $-0.71\pm   0.05$ &   B07  \\
NGC\,4013 & $0.151\pm  0.015$  &1575  & $0.0378\pm  0.0008$ & $-0.58\pm   0.04$ &   W15  \\
NGC\,4096 & $0.198\pm  0.020$  &1575  & $0.0560\pm  0.0011$ & $-0.53\pm   0.04$ &   W15  \\
NGC\,4125 & $0.012\pm  0.001$  &1365  & $0.0018\pm  0.0002$ & $-0.84\pm   0.06$ &   B07  \\
NGC\,4157 & $1.021\pm  0.102$  &1575  & $0.1845\pm  0.0037$ & $-0.72\pm   0.04$ &   W15  \\
NGC\,4214 & $0.126\pm  0.013$  &1400  & $0.0620\pm  0.0050$ & $-0.31\pm   0.06$ &   Y01  \\  
NGC\,4217 & $0.446\pm  0.045$  &1575  & $0.1116\pm  0.0022$ & $-0.58\pm   0.04$ &   W15  \\
NGC\,4236	 & $0.037 \pm  0.004$  & 1365 & $0.0171\pm  0.0009$ & $-0.35\pm   0.05$ & B07     \\
NGC\,4244 & $0.050\pm  0.005$  &1575  & $0.0165\pm  0.0006$ & $-0.46\pm   0.04$ &   W15  \\
NGC\,4254	 & $2.556 \pm  0.256$  & 1365 & $0.4300\pm  0.0215$ & $-0.79\pm   0.05$ & B07     \\
NGC\,4321	 & $1.520 \pm  0.152$  & 1365 & $0.3100\pm  0.0100$ & $-0.71\pm   0.05$ & B07     \\
NGC\,4449 & $1.032\pm  0.103$  &1400  & $0.3000\pm  0.0300$ & $-0.54\pm   0.06$ &   NVSS \\  
NGC\,4450	 & $0.041 \pm  0.004$  & 1365 & $0.0130\pm  0.0020$ & $-0.51\pm   0.08$ & B07     \\
NGC\,4559 & $0.236\pm  0.024$  &1365  & $0.0810\pm  0.0100$ & $-0.48\pm   0.07$ &   B07  \\
NGC\,4565	 & $0.968 \pm  0.097$  & 1575 & $0.1522\pm  0.0030$ & $-0.77\pm   0.04$ & W15   \\
NGC\,4631 & $4.640\pm  0.464$  &1575  & $1.1340\pm  0.0370$ & $-0.59\pm   0.04$ &   W15  \\
NGC\,4725 & $0.220\pm  0.022$  &1365  & $0.0540\pm  0.0050$ & $-0.62\pm   0.06$ &   B07  \\
NGC\,4736 & $0.897\pm  0.090$  &1365  & $0.3100\pm  0.0200$ & $-0.47\pm   0.05$ &   B07  \\
NGC\,4826	 & $0.247 \pm  0.025$  & 1365 & $0.1100\pm  0.0100$ & $-0.36\pm   0.06$ & B07     \\
NGC\,5033 & $1.144\pm  0.114$  &1365  & $0.2200\pm  0.0100$ & $-0.73\pm   0.05$ &   B07  \\
\hline                                                                                                   
\end{tabular}                                                                                            
\end{table*}

\setcounter{table}{2}
\begin{table*}
\centering
\caption{Radio spectral indices between 144 and approximately 1400\,MHz.}
\label{tab:spix2}
\begin{tabular}{l ccccc}
\hline\hline
Galaxy  & $S_{144}$ & $\nu$ & $S_\nu$  &  $\alpha$   &  Ref. \\
& (Jy)               & (MHz) & (Jy) & & \\\hline
NGC\,5055 & $2.286\pm  0.229$  &1365  & $0.4200\pm  0.0100$ & $-0.75\pm   0.05$ &   B07  \\
NGC\,5194 & $6.363\pm  0.636$  &1365  & $1.4100\pm  0.0100$ & $-0.67\pm   0.04$ &   B07  \\
NGC\,5195 & $1.004\pm  0.100$  &1365  & $0.2000\pm  0.0200$ & $-0.72\pm   0.06$ &   B07  \\
NGC\,5297 & $0.118\pm  0.012$  &1575  & $0.0238\pm  0.0005$ & $-0.67\pm   0.04$ &   W15  \\
NGC\,5457 & $3.226\pm  0.323$  &1400  & $0.8080\pm  0.0800$ & $-0.61\pm   0.06$ &   W92 \\  
NGC\,5474 & $0.021\pm  0.002$  &1400  & $0.0120\pm  0.0010$ & $-0.25\pm   0.06$ &   NVSS \\  
NGC\,5775	 & $1.212 \pm  0.121$  & 1575 & $0.2550\pm  0.0051$ & $-0.65\pm   0.04$ & W15   \\  
NGC\,5866 & $0.065\pm  0.006$  &1400  & $0.0230\pm  0.0020$ & $-0.46\pm   0.06$ &   Y01  \\  
NGC\,5907 & $0.594\pm  0.059$  &1575  & $0.1187\pm  0.0036$ & $-0.67\pm   0.04$ &   W15  \\
NGC\,6946  & $6.578 \pm  0.658$  & 1365 & $1.7000\pm  0.0100$ & $-0.60\pm   0.04$ & B07     \\
NGC\,7331 & $2.122\pm  0.212$  &1365  & $0.4200\pm  0.0200$ & $-0.72\pm   0.05$ &   B07  \\
\hline                                                                                                   
\end{tabular}                                                                                            
\tablefoot{References: A13: \citet{adebahr_13a}; B07:
  \citet{braun_07a}; B17: \citet{brown_17a}; H:18: \citet{heesen_18a};
  NVSS: integrated from NRAO Sky Survey maps \citep{condon_98a};
  T07: \citet{tabatabaei_07a}; T17: \citet{tabatabaei_17a}; W15:
  \citet{wiegert_15a}; W92: \citet{white_92a}; Y01: \citet{yun_01a}.}
\end{table*}

\end{document}